\documentclass[final,1p,times]{elsarticle}
\usepackage{amssymb}
\usepackage{amsmath}
\usepackage{amsthm}

\usepackage{color}%

\begin{document}

\begin{frontmatter}



\title{Kinetic inflation in deformed phase space Brans-Dicke cosmology}


\author[1,2]{S. M. M. Rasouli}
\author[1,2]{J. Marto}
\author[1,2,3]{P. V. Moniz}

\address[1]{Departamento de F\'{i}sica,
Universidade da Beira Interior, Rua Marqu\^{e}s d'Avila
e Bolama, 6200-001 Covilh\~{a}, Portugal.}
\address[2]{Centro de Matem\'{a}tica e Aplica\c{c}\~{o}es (CMA - UBI),
Universidade da Beira Interior, Rua Marqu\^{e}s d'Avila
e Bolama, 6200-001 Covilh\~{a}, Portugal.}
\address[3]{DAMTP, Centre for Mathematical Sciences, University of Cambridge, Wilberforce Road, Cambridge CB3 0WA, UK}

\begin{abstract}
In this paper,
by establishing a Brans-Dicke (BD) cosmology by means of a deformed phase space, in the
absence of any scalar potential, cosmological constant and ordinary matter, we show
that it is feasible to overcome obstacles reported in the corresponding
commutative (non-deformed) frameworks. More concretely, by applying the Hamiltonian
formalism and introducing a dynamical deformation, between the momenta
associated to the FLRW scale factor and the BD scalar field, we obtain the
modified equations of motion. In particular, these equations reduce to their
standard counterparts when the noncommutative (NC) parameter is switched off.
By focusing on a specific branch of solutions, in contrast to
standard frameworks (even with a varying BD coupling parameter), we show that we can obtain an adequate
appropriate inflationary epoch possessing a suitable graceful exit.
In other words, in the Jordan frame (JF), such branch of solutions properly satisfy
the sufficient condition required for satisfactory inflation, which is equivalent to
get an inflationary phase in the conformal Einstein
frame (EF) without branch change.
%Moreover, in the EF, by moving to the conformally
%flat background, it might be expected that our NC equations of
%motion to be correspond to the quantum BD cosmological model established for the spinor case.
Concerning the cosmological dynamics, we further show that our NC
framework bears close resemblance to the $R^2$ (Starobinsky) inflationary model.
%but also the phase space
%portrait authenticates the above mentioned achievements associated to the inflationary phase. Furthermore,
%the NC effects show themselves by removing the big bang singularity
 %at the early times as well as by yielding the constant scale factor at late times.  }
\end{abstract}

\begin{keyword}
inflation \sep noncommutativity \sep Brans-Dicke theory
%% \PACS 02.40.Gh \sep 98.80.-K \sep 98.80.Cq

\end{keyword}

\end{frontmatter}


\section{Introduction}

%As an alternative theory to the Einstein general relativity, the BD theory \cite{}, the simplest version of the scalar tensor theories \cite{}, has been %widely applied to describe various epoches of the universe \cite{}. In the action of the standard BD theory~\cite{}, the dynamical BD free scalar field %plays the role of inverse of the Newton gravitation constant. Moreover, there is a dimensionless coupling parameter which is constant in the standard BD %theory.

One of the strongest motivations, in cosmology, for applying various scenarios of scalar tensor theories, including BD
theory~\cite{BD61} as the simplest prototype, has been to overcome the problems associated to standard
 inflationary scenarios \cite{LS89, W89, SA90, BM90, NOO15}. In these applications as well as for constructing
 quintessential scenarios (see, e.g., \cite{T02, Faraoni.book} and related papers) associated to the present accelerating epoch,
 a scalar potential is usually added by hand into the action of the standard BD
 theory\rlap.\footnote{Recently, concerning such an ad hoc scalar potential in scalar tensor theories,
 it has been shown that the geometry of the higher dimensions can instead naturally yield it; see, for instance,~\cite{RFS11,RFM14,RM16, RM18}. Moreover, we should note that, in addition to including a scalar potential, there
 are also other approaches for describing the present day acceleration (which, in turn,
 suffer from major problems \cite{Ponce1}): (i) When the BD coupling parameter is
 restricted to $-2\leq \omega\leq -3/2$; in this case, the energy
 condition associated to the BD scalar field is not satisfied.
 In this paper, to avoid encountering such a problem, we
 always assume $-3/2< \omega$. (ii) Assuming a varying BD
 coupling parameter \cite{BP01}. (iii) Assuming that the scalar
 fields can interact with ordinary matter, which are called
 chameleon fields \cite{DB08}. (iv) By starting from the BD theory in five
 or higher dimensional space-times, see for instance, \cite{QMHY05}. We should emphasize that
 the motivation of this paper is to advance a novel manner to address and contribute towards the problems
 associated to the conventional kinetic inflation (of the early universe) rather than the
 present accelerating universe. Nevertheless, we show that our chosen noncommutativity affects also on the behavior of the scale factor of the
 late universe. It is important to note that it might be feasible to describe
 the present day acceleration by employing either the deformed Poisson
 bracket (which is introduced in this paper) or different kinds of noncommutativity.
 However, in this paper, we merely focus our attention on the kinetic inflationary phase of the early universe.}
 Moreover, it should be noted that
 inflationary scenarios have also been presented by establishing
 quantum BD cosmology, see, for instance, \cite{HCC95,CHC95,GOZ99,GK00}.

Nevertheless, in contrast to that standard inflationary
strategy, it has also been considered that the accelerating scale factor
associated to such epoch can be obtained merely from
the dynamics of the scalar field, without employing either
the corresponding scalar potential or the cosmological constant~\cite{GSV91,Lev95,Lev95-2, RM14,RM-Odes}.
However, it has been demonstrated that such kinetic inflationary scenarios, as investigated in the BD theory
(even with varying BD coupling parameter), suffer
from a pertinent set of problems~\cite{Lev95-2}.
More concretely, for the inflating scale factor
 associated to all the D-branch\footnote{There are two different branches of cosmological
 solutions in the standard BD theory, which are called the D- and X-branches.
 Following \cite{Lev95, RM14}, let us describe them as follows.
In the conformal EF, the scale
factor associated to the X-branch and D-branch always
 expands and contracts, respectively. Although, in the JF, their behaviors would
 effectively depend on the values taken by
 the BD coupling parameter, nevertheless, these solutions correspond to those expand and
 contract in the EF are also referred as X- and D-branch solutions, respectively
 (see also the explanations below equation (\ref{g-f}) in section \ref{Vacuum-NC-BD}, subsection \ref{Standard case} and the paragraph including equation (\ref{a-sec}).}).
solutions in the JF, there is no successful transition
between this accelerating epoch and the decelerating
expansion, which is known as the graceful exit problem\rlap.\footnote{Moreover, the accelerating epoch obtained in the context
 of string theory \cite{BV94}, driven by an inflaton field
 under the influence of a potential, also encounters the same problem. It has been argued that this
 problem is a consequence of a cosmological singularity being reached in a finite proper time
 \cite{LM00}. It has been demonstrated that the singularity can be removed by including quantum effects \cite{L97,GMV96}.}
   Furthermore, satisfying a sufficient condition for inflation
 imposes the scale factor of EF to accelerate.
 However, concerning a kinetic inflationary model in such generalized BD theory, there is no
  source to yield such an acceleration stage~\cite{Lev95-2}. In this paper we address that obstacle
in the context of the BD theory from the perspective of a NC setting and show
that it is feasible to overcome such a problem.

  For the X-branch, the situation is instead quite different. Let us be more precise.
 In the absence of any scalar potential and ordinary matter in the standard BD theory
 (where the BD coupling parameter is constant), the only class of solutions
 which can yield an accelerating scale factor is the D-branch solutions, with
 negative values for both the BD coupling parameter
 and the time derivative of the scalar field. The case
 where the BD coupling parameter is equal to $-1$, have been investigated in \cite{V91,GV93}.
However, it is impossible to get a gravity-driven acceleration
for the X-branch solutions in the strict context of the standard BD theory for the spatially flat FLRW background.
One viable alternative to overcome this situation and obtaining
gravity-driven acceleration in the X-branch, is applying a generalized
BD theory where the BD coupling parameter can vary, see, e.g.,~\cite{Lev95-2}.
However, these models have, in turn, their shortcomings.
For instance, as argued in~\cite{Lev95-2}, similarly to the
D-branch solutions, we find a lack of accelerating scale
factor in the EF, or equivalently, the graceful exit problem remains.

At high energy regimes,
%are the subject matter of our discussions, it would seem that such required sources, which can assist to
 to overcome the mentioned obstacles, features as motivated from tentative quantum gravity proposals, have been considered.
%can be emerged from the quantum effects.
%In order to explain the substantive behavior of the physical quantities
%in very small scales, we should employ a quantum theory.
%For instance, to describe the universe near the big bang, where the
%quantum effects are substantial, it has been believed
%that the background gravitational theory must be incorporated with the
%quantum theory to construct a viable quantum gravity setting.
%In this respect, it has been undertaken a lot of endeavors to construct a theory of quantum gravity, see,
%e.g., \cite{GSW88,Polchinski.book98,Rovelli.book04,Thiemann.book07}.
Among such physical
frameworks, a deformation in the phase space structure \cite{ZFC.book05}, emerging via a length parameter
[which is usually interpreted as the Planck (length) constant] to the
background theory, has been considered as a promising approach.
It should be emphasized that the standard results
are recovered in a proper limit of the length parameter.
The implications associated to these frameworks in cosmology \cite{GOR02,BP04} have included the
removal of the big bang singularity at early times as well as coarse
grained effects at late times of the evolution of the universe.
Recently, various NC cosmological/gravitational frameworks have been constructed to
investigate different shortcomings associated to the
corresponding standard models~\cite{RFK11,GSS11, RZMM14,RZJM16,LSR18,RSFMM18}. For instance, in \cite{RM14}, two of us have
 shown that such source can be borrowed from the NC portions accompanying the standard BD setting.
 Therefore, by focusing on the D-branch solutions, the graceful exit problem has been solved appropriately.
%As it has been believed that the
NC phase space models
have also been  proposed within general settings emerging from a generalized
uncertainty principle,
%it is worthwhile to refers such investigation in cosmology,
see, for instance \cite{JRM14,GWY16,SHFA17}.

In the context of the above paragraphs, the main objective of the present paper is to show that
by applying a BD cosmology and by employing a dynamical deformation in the phase space,
it is feasible to overcome the obstacles associated to the X-branch solutions.
Moreover, we show that our herein NC model has other salient features which can be related
to other interesting inflationary models constructed in either
BD theory or general relativity by including quantum corrections.
Specifically, (i) the big bang singularity
is removed; (ii) an accelerating inflationary phase is obtained for
early times, which has a graceful exit; (iii) for the late times, the scale
factor is in a zero acceleration phase; (iv) the solutions fully satisfy
the nominal and sufficient conditions associated to inflation;
(v) the cosmological phase space analysis not only confirms the X-branch to be
successful inflation but also provides situations for plausible
resemblance to the Starobinsky inflationary model \cite{S80}.
Furthermore, by analyzing the time behaviors associated to either scale
factor in the conformal EF or its equivalent quantity in the JF, we
 show that our herein NC framework can provide a viable scenario for a successful inflation in both of these frames.
 In addition, deriving the EF equations of motion in
 a conformably flat background motivates to find
 definite conditions (at least by means of numerical methods) to obtain a possible
 correspondence between our NC model and the quantum BD cosmological
 framework in the EF, in which the scalar field is coupled to Dirac spinors~\cite{Faraoni.book}.

The paper is briefly outlined as follows.
In the next section, the cosmological equations of motion for the
generalized BD cosmology, in the presence a general dynamical
deformation (noncommutativity), will be obtained. In section \ref{Vacuum-NC-BD}, we
first present solutions of the equations of motion for the non-deformed (commutative) case,
analytically. Subsequently, by specifying the deformed Poisson bracket, we
show that the evolutionary equations depend only on the dynamics of the BD scalar field. Then, we solve analytically the NC equations
for particular cases. For the general NC case, we apply numerical
methods to depict the time behavior of physical quantities. In section \ref{Horizon},
we show that the nominal as well as the sufficient conditions are completely satisfied
for the solutions associated to the inflation obtained in \ref{Vacuum-NC-BD}.
In section \ref{dynamical}, we compare the evolutionary equation for
the scale factor associated to our herein model with that of the Starobinsky model.
Subsequently, by depicting the phase space portrait, we show that the results of the
section \ref{Vacuum-NC-BD} for the X-branch solutions are reconfirmed. In section \ref{EF}, we proceed to derive
the results in the conformal EF and present complementary comments.
Finally, in section \ref{conclusion}, we summarize the main results
of the work and present additional discussions.
%In \ref{AppA}, a few figures are
 %provided, strengthening the argument towards validity of our NC framework
 %for new appropriate sets of ICs and present parameters of the model.

\section{Deformed phase space Brans-Dicke setting}
\indent
\label{NC-BDT}
%In this section, we review the BD theory in the Hamiltonian
%formalism.
We consider a background geometry which is described with the spatially flat Friedmann-Lema\^{\i}tre-Robertson-Walker~(FLRW)
line-element as
\begin{equation}\label{metric1}
ds^{2}=-\mathcal{N}^2(t)dt^2+a(t)^{2}\left(dx^2+dy^2+dz^2\right),
\end{equation}
where $a(t)$ is the scale factor, $\mathcal{N}(t)$ is a lapse function and $(x,y,z)$ denote the Cartesian coordinates.
In the JF, the Lagrangian density associated to the scalar
tensor gravity can be written as
\begin{eqnarray}
{\cal L}[\gamma,\phi]=\sqrt{-\gamma}\left[\phi R-
\frac{\omega(\phi)}{\phi}\gamma^{\mu\nu}\nabla_\mu\phi\nabla_\nu\phi-V(\phi)\right],\label{lag1}
\end{eqnarray}
where the Greek indices run from zero to $3$, $\gamma$ and $R$ stand for the determinant
and the Ricci scalar associated to the metric $\gamma_{\mu\nu}$,
respectively;
%${\cal L_{\rm matt}}=16\pi\rho(a)$ [where $\rho(a)$ is the energy density]
%is the Lagrangian density associated to ordinary matter;
$\phi$ is the BD scalar field (which is normalized to $G^{-1}$ where $G$ is Newton gravitational constant)
such that, as throughout this paper, we
will use the Planck units, its present value
should be equal to unity~\cite{BM90}; $V(\phi)$ is the scalar potential.
Moreover, for attractive gravity, we must assume $\phi>0$.

%%%%%%%%%%%%%%%%%%%%%%%%%%%%%%%%%%%%%%%%%%%%%%%%%%%%%%%%%%%%%%%%%%%%%%%%%%%%%%%%%%%%%%%%%%%%%%%%%%%
%%%%%%%%%%%%%%%%%%%%%%%%%%%%%%%%%%%%%%%%%%%%%%%%%%%%%%%%%%%%%%%%%%%%%%%%%%%%%%%%%%%%%%%%%%%%%%%%
%%%%%%%%%%%%%%%%%%%%%%%%%%%%%%%%%%%%%%%%%%%%%%%%%%%%%%%%%%%%%%%%%%%%%%%%%%%%%%%%%%%%%%%%%%%%%%%%

In a particular case where $\omega(\phi)=\omega_0={\rm constant}$, the Lagrangian~(\ref{lag1})
reduces to the Lagrangian of the BD theory in the presence of a scalar potential.
In the next sections, for the sake of simplicity in analyzing cosmological
solutions in the presence of the NC effects, not only we
work within standard BD theory (namely, constant values of the coupling parameter)
but also we will assume the very simple case where the scalar potential is absent.

%for non--ghost scalar field
%situations~\cite{BKM04,DDB07,BS07,B09}.
It is easy to show that the Hamiltonian associated to (\ref{lag1}) can be written as
\begin{eqnarray}\label{Ham-1}
{\cal H}=\frac{\mathcal{N}}{6\xi^2}
\left[-\frac{\omega(\phi)}{6a\phi}P_a^2+a^{-3}\phi P_\phi^2-a^{-2}P_a P_\phi\right]+\mathcal{N}a^3V(\phi),
\end{eqnarray}
 where
 \begin{eqnarray}\label{xi}
\xi\equiv\left[1+\frac{2\omega(\phi)}{3}\right]^\frac{1}{2}\label{zi},
\end{eqnarray}
and $P_a$ and $P_\phi$ stand for the
conjugate momenta associated to the scale factor and scalar field, respectively.

%the spatial gradients in
%the BD scalar field are negligible, namely,
% By using the equation of state associated
%to a perfect fluid and the Hamiltonian constraint,
%it is straightforward to derive the usual FLRW field equations in the context of the BD cosmology.
%However, in this paper, we prefer to work with the first order Hamiltonian differential equations.
%As the BD equations of motion for the above model
%have been given in the literature, let us before deriving them
%%%%%%%%%%%%%%%%%%%%%%%%%%%%%%%%%%%%%%%%%%%%%%%%%%%%%%%%%%%%%%%%%%%%%%%%%%%%%%%%%%%%%%%
%%%%%%%%%%%%%%%%%%%%%%%%%%%%%%%%%%%%%%%%%%%%%%%%%%%%%%%%%%%%%%%%%%%%%%%%%%%%%%%%%%%%%%%%%%%%%
%%%%%%%%%%%%%%%%%%%%%%%%%%%%%%%%%%%%%%%%%%%%%%%%%%%%%%%%%%%%%%%%%%%%%%%%%%%%%%%%%%%%%%%%%%%

In order to obtain and investigate an acceleration of the universe, solely driven from the kinematics,
%gravity and investigating the corresponding inflationary universe,
which will be studied in the next sections, let us assume that
 %the Lagrangian density of the ordinary matter as well as
 the scalar potential is absent.
 Instead, we extend the standard model by investigating a generalized setting in which the
 Poisson bracket associated to the conjugate momenta of the scale factor and
 the BD scalar field does not vanish. More precisely, we will
  generalize the noncommutativity proposed in~\cite{RFK11}.
  It is straightforward to show that such a dynamical NC model can be constructed based upon a dimensional analysis~\cite{RFK11, RZMM14, RSFMM18}.
%We will show that such NC model yields interesting dynamics
%such that it is feasible to overcome to some shortcomings associated to the %corresponding standard frameworks.

In this work, let us propose the Poisson commutation relation between
the momenta as\footnote{Such a choice of dynamical deformation between the conjugate momentum sector might be
corroborated by the physical arguments presented
in \cite{RFK11,RZMM14,RSFMM18}, see also \cite{FGMG17}. Let us be more precise.
Motivation for employing such type of noncommutativity, following \cite{RFK11}, can be outlined as follows.
(i) A few clues might be given in investigations based on string theory, for
instance, in the flux compactification, as referred in \cite{KSS07}. (ii) The time dependent dynamical deformations in the
phase space have also been chosen in frameworks motivated by
generalized uncertainty principle \cite{KMM95,KM97} and $\kappa$-Minkowskian
space-time \cite{KGN02,KG05}, which have been applied in cosmological phase space \cite{KS08}.
(iii) Noncommutativity between momentum conjugate bears close resemblance
with the behaviors observed from charged particles in the presence of a magnetic field \cite{MF14}.
(iv) In general, it can be considered as an extension of the usual noncommutativity between the spatial coordinates \cite{DS04}.
In addition to the above mentioned
justifications, the chosen noncommutativity in our
herein model has been motivated and supported from
dimensionality analysis.
%{\cob: assuming the scale factor and the lapse
%function as dimensionless variables, $[x^\mu]=L$ and $\hbar=1=c$, then according to
%the Plank length, $L_{\rm P}=\sqrt{\hbar G/c^3}$, it
%is straightforward to show that $[G]=L^2$, $[\phi]=L^{-2}$, $[P_a]=L^{-3}$,
%$[P_\phi]=L^{-1}$, and $[\{P_a,P_\phi\}]=L^{-1}$.
%Therefore, choosing relation~(5) yields $[\theta]=L$.}
Furthermore, in the previous investigations, it has been shown that
employing such a particular kind of deformations leads
some appropriate results, see, e.g., \cite{RZMM14,RSFMM18}.}
%\cite{RFK11}.Moreover, the interesting physical results produced due to the proposing such a
%deformations
  %in turn, might protect constructing such toy models.}}
\begin{eqnarray}
 \{P_a,P_\phi\}=\theta \zeta\phi,\label{NC-Poisson}
\end{eqnarray}
where $\theta$ is a constant NC parameter and $\zeta=\zeta(a)$ is an arbitrary function of the scale factor;
we leave the Poisson brackets associated to the other variables unchanged, namely
\begin{eqnarray}
\{a,\phi\}=0,\hspace{8mm}\{\phi,P_\phi\}=1, \hspace{8mm}\{a,P_a\}=1.\label{NC-Poisson-2}
\end{eqnarray}

Furthermore, we will work in the comoving gauge; namely, we would like to use the cosmic time and therefore we should set $\mathcal{N}(t)=1$.
Moreover, we begin in obtaining the general formula associated to the generalized BD theory in the absence of the scalar potential, in which the dynamical Poisson bracket (\ref{NC-Poisson}) is used.
Subsequently, in the next section we will employ a constant BD coupling parameter.

By employing the commutation relations~(\ref{NC-Poisson}) and (\ref{NC-Poisson-2}), the modified equations of motion
associated to the Hamiltonian~(\ref{Ham-1}) are given by
\begin{eqnarray}
\dot{a}\!&=&\!-\frac{1}{6\xi^2a}
\left[\frac{\omega}{3\phi}P_a+a^{-1} P_\phi\right],\label{NC-a}\\
\dot{P_a}\!&=&\!-\frac{1}{36\xi^2a^4\phi}
\left[\omega a^2P^2_a+12a\phi P_a P_\phi-18\phi^2P_\phi^2\right]
+\frac{\theta\zeta(a)\phi}{6\xi^2a^3}\left(2\phi P_\phi-aP_a\right),\label{NC-Pa}\\
 \dot{\phi}\!&=&\!-\frac{1}{6\xi^2a}
\left(a^{-1}P_a-2a^{-2}\phi P_\phi\right),\label{NC-phi}\\\nonumber
\\
\dot{P_\phi}\!&=&\!\frac{-1}{36\xi^2a^3\phi^2}
\left[\omega a^2P^2_a+6\phi^2P_\phi^2\right]+\frac{P_a^2}{36\xi^2a\phi}\frac{d\omega(\phi)}{d\phi}
+\frac{\theta\zeta(a)}{18\xi^2a^2}\left[3\phi P_\phi+\omega(\phi)aP_a\right]
\label{NC-Pphi},
\end{eqnarray}
where an overdot stands for the differentiation with respect to the cosmic time.
Moreover, as we have considered the homogeneous and isotropic FLRW
universe, we have assumed that the BD scalar field to be merely a function of the cosmic time.
Additionally, we should notice that, by choosing the noncommutativity (\ref{NC-Poisson}), just
the equations of motion associated to the conjugate momenta, i.e.,~(\ref{NC-Pa}) and (\ref{NC-Pphi}), have been modified.
It is important to note that when $\theta=0$,
equations (\ref{NC-a})-(\ref{NC-Pphi}) reduce to those of the corresponding
commutative (non-deformed) counterparts.

 After some manipulations, it is straightforward to
 show that the NC equations can be written, as the modified versions of the standard form, as
  \begin{eqnarray}\label{H2}
H^2\!&=&\!\frac{\omega}{6}\left(\frac{\dot{\phi}}{\phi}\right)^2-H\left(\frac{\dot{\phi}}{\phi}\right),\\\nonumber
\\
\frac{\ddot{a}}{a}\!&=&\!-\frac{\omega}{3}\left(\frac{\dot{\phi}}{\phi}\right)^2
+H\left(\frac{\dot{\phi}}{\phi}\right)+\frac{1}{6\xi^2\phi}\frac{d\omega(\phi)}{d\phi}\left(\dot{\phi}\right)^2
+\frac{\theta\zeta(a)\phi}{18\xi^2a^2}\left(3H-\frac{\omega\dot{\phi}}{\phi}\right)
,\label{a2dot}\\\nonumber
\\
 \ddot{\phi}\!&+&\!3H\dot{\phi}=-\frac{1}{3\xi^2}\frac{d\omega(\phi)}{d\phi}\left(\dot{\phi}\right)^2
 -\frac{\theta\zeta(a)\phi^2}{6\xi^2a^2}\left(2H+\frac{\dot{\phi}}{\phi}\right),\label{phi2dot}
\end{eqnarray}
where $H=\dot{a}/a$ is the Hubble parameter.
It is important to note that the Hamiltonian constraint was not modified
under the dynamical deformation between the momenta, which leads to a power-law
relation between the scale factor and the scalar field as in the
 corresponding case of the standard BD theory~\cite{Lev95,Lev95-2}.
 However, the acceleration equation, as well
 as the wave equation (associated to the BD scalar), have been
 modified such that in the absence of the NC parameter we recover
 the corresponding commutative equations~\cite{Lev95,Lev95-2}.
We should point that we have
 three nonlinear differential equations
 (which are not independent) with two unknowns\footnote{Note that $\zeta=\zeta(a)$ is not another
 independent dynamical variable; indeed, we will proceed to
 get the solutions by specifying this function.} $a$ and $\phi$.
 In this paper, the extra terms, which were produced due to the
 deformation in the phase space, will mostly require our attention.

\section{Brans-Dicke cosmological vacuum solutions in deformed phase space }
\indent
\label{Vacuum-NC-BD}
Let us focus on the equations (\ref{H2})-(\ref{phi2dot}) and work
in the context of standard BD theory as the background theory\rlap.\footnote{Hereafter, we always set $\omega=\omega_0={\rm constant}$.
For simplicity, let us drop the index $0$ from $\omega_0$.}
We should refer to the gravity driven acceleration and kinetic inflation
investigated in non-deformed cases of the generalized BD theory
[namely, solutions of equations (\ref{H2})-(\ref{phi2dot}) by assuming $\theta=0$]
in interesting works by Levin~\cite{Lev95,Lev95-2}.
In our work, we would like to investigate instead how effects imported
from deformation in the phase space can overcome the problems of the non-deformed cases.

In analogy with the standard BD theory, we should determine the energy density
and the pressure associated to the $\phi$-field, e.g., \cite{Lev95}
 \begin{eqnarray}\label{ro-1}
 H^2\!\!&=&\!\!\frac{8\pi}{3\phi}\rho_\phi,\\
 \frac{\ddot{a}}{a}\!\!&=&\!\!-\frac{4\pi}{3\phi}(\rho_\phi+3p_\phi).\label{pr-1}
 \end{eqnarray}
By comparing equations (\ref{H2}) and (\ref{a2dot}) with (\ref{ro-1}) and (\ref{pr-1}), respectively,
for a constant $\omega$, we retrieve
 \begin{eqnarray}\label{ro-2}
\rho_\phi&=&\frac{\omega}{16\pi}\frac{(\dot{\phi})^2}{\phi}-\frac{3}{8\pi}H\dot{\phi},
\\\nonumber
\\
p_\phi&=&\frac{\omega}{16\pi}\frac{(\dot{\phi})^2}{\phi}-\frac{1}{8\pi}H\dot{\phi}
-\frac{\theta\zeta(a)}{24\pi\xi^2}\left(\frac{\phi}{a}\right)^2\left(H-\frac{\omega}{3}\frac{\dot{\phi}}{\phi}\right).\label{pr-2}
\end{eqnarray}

Let us proceed our investigation by analyzing the solutions associated to field
equations (\ref{H2})-(\ref{phi2dot}) for both the commutative and NC cases for constant $\omega$.
 From (\ref{H2}), we obtain
 \begin{eqnarray}\label{H-a}
 H=\chi\left(\frac{\dot{\phi}}{\phi}\right), \hspace{10mm}a(t)=a_i\left[\phi(t)\right]^\chi,
 \end{eqnarray}
where $a_i$ is an integration constant and\footnote{Hereafter, situations where
$\chi$ and $\lambda$ have no explicit index, it means that we discuss both the signs $+$ and $-$.}
\begin{eqnarray}\label{g-f}
\chi=\chi_{\pm}\equiv-\frac{1}{2}\left(1+\lambda \xi\right), \hspace{10mm}\lambda=\lambda_{\pm}=\pm1.
 \end{eqnarray}
In this paper, $\lambda_+$ and $\lambda_-$ are associated to what we will
denominate as the upper (plus sign) and the lower (minus sign) solutions.
Here, in analogy with the non-deformed BD theory
(either with constant or varying $\omega$), we should distinguish the cosmological solutions.
It has been shown that in the EF, there are
two different types of the cosmological solutions, corresponding to
different values of $\lambda$ (see section \ref{EF}): in one type, the
 universe contracts while in the other one, it expands, see, e.g., \cite{Lev95}.
However, in the JF, for either of the signs, we can still get an expanding
universe [cf discussions below equation (\ref{OT-a2}), in which
we will show under which conditions, we
can get expanding universe in the JF for both of the cases].
Since for $\theta=0$, our NC solutions should always
 reduce to the standard case, following \cite{Lev95, RM14}, let us then distinguish these solutions as follows.
 In our herein NC model, the $X$ and $D$ branches correspond to the modified solutions
 whose scale factor expands and contracts in the EF, respectively, for the particular case where $\theta=0$ (see also footnote 2).
 Consequently, throughout this paper, the X and D branches correspond to the upper and lower signs, respectively.

According to relations (\ref{H-a}) and (\ref{g-f}), we obtain two branches for the Hubble parameter $H$:
 (i) By assuming $\omega<0$ ($\xi<1$), for both of the branches (namely, for both values of $\lambda$),
 we obtain contracting and expanding universe provided that $\dot{\phi}>0$ and $\dot{\phi}<0$, respectively.
 (ii) Letting $\omega>0$ ($\xi>1$), we obtain $H>0$, if we choose either the upper sign together with $\dot{\phi}<0$ or
 the lower sign together with $\dot{\phi}>0$.
We should note that due to the presence of the NC parameter
in the field equations, it affects the dynamics
of the variables, and, as it will be shown in the next
sections, so it is not easy to anticipate the time behavior of
the quantities.

In what follows, we will show that it is possible to write both $\dot{\phi}$ and
$\ddot{a}/a$ (in the corresponding differential equations) as functions of the BD scalar field.
It is easy to show that equation (\ref{phi2dot}) yields
\begin{eqnarray}\label{phidot-1}
a^3\dot{\phi}+b=-\frac{\theta(2\chi+1)}{12\xi^2}\int a\zeta(a)d(\phi^2),
 \end{eqnarray}
where $b$ is an integration constant, which can be positive, negative or zero.
Using integration by parts and employing (\ref{H-a}) into the retrieved equation, we obtain
\begin{eqnarray}
\dot{\phi}=-\frac{b}{a_i^3 \phi^{3\chi}}-\left[\frac{\theta(2\chi+1)}{12a_i^2\xi^2}\right]\left[\phi^{2(1-\chi)}\zeta(a)-\frac{1}{a_i\phi^{3\chi}}\int \phi^2d\left[a\zeta(a)\right]\right]\equiv f(\phi).\label{phidot-2}
 \end{eqnarray}
From relation (\ref{H-a}), it is seen that the scale factor is a function
of the BD scalar field, therefore, the right hand side of (\ref{phidot-2}) is
just a function of $\phi(t)$, namely $\dot{\phi}=f(\phi)$.

Moreover, it is straightforward to show that the quantity $\ddot{a}/a$ can
also be written merely as a function of the BD scalar field, as
\begin{eqnarray}\label{a2dot-2}
\frac{\ddot{a}}{a}=
\left[1+\frac{\theta \zeta(a)\phi^{\chi+2}}{6a_i^2\xi^2 f(\phi)}\right]\left[\frac{f(\phi)}{\phi}\right]^2,
\end{eqnarray}
where we have used (\ref{H-a}) and $\dot{\phi}=f(\phi)$ is given by (\ref{phidot-2}).

\subsection{Standard case: $\theta=0$}
\label{Standard case}
In order to compare the solutions
 associated to the deformed case with those obtained in the standard case, let us
  review the results associated to the standard BD theory for the spatially flat
  FLRW metric in vacuum which are known as O'Hanlon-Tupper
  solutions~\cite{o'hanlon-tupper-72-KE95-MW95, C98}. Substituting $\theta=0$ in (\ref{phidot-2})
  and using (\ref{H-a}), we obtain the following solutions
\begin{eqnarray}\label{OT-phi1}
\phi(t)&=&\left[C(t-t_i)\right]^{\frac{1}{3\chi+1}},\\
a(t)&=&a_i\left[C(t-t_i)\right]^{\frac{\chi}{3\chi+1}},\label{OT-a1}
\end{eqnarray}
 for $\chi\neq-1/3$ ($\omega\neq-\frac{4}{3}$), where $C\equiv-\frac{b(3\chi+1)}{a_i^3}$
and
\begin{eqnarray}\label{OT-phi2}
\phi(t)&=&\phi_i{\rm e}^{-\frac{c}{a_i}(t-t_i)},\\
a(t)&=&a_i\phi_i^{-\frac{1}{3}}{\rm e}^{\frac{c}{3a_i}(t-t_i)},\label{OT-a2}
\end{eqnarray}
 for $\chi=-1/3$ ($\omega=-\frac{4}{3}$), where $\phi_i$ and $t_i$ are integration constants.

Let us analyze the time behaviors of the BD scalar field and the scale
factor in solutions (\ref{OT-phi1}) and (\ref{OT-a1}) by
considering different conditions for the integration constants and the BD coupling parameter.
More concretely, for $C>0$, $\omega>-4/3$ and $t>t_i$, we obtain two different sets of the solutions.
 (i) For $\lambda_-=-1$, from the progressing cosmic time, the
BD field always increases, while $G_{\it eff}\simeq1/\phi$ and the scale factor
always contract. (ii) For $\lambda_+=1$, as cosmic time increases, the BD field
decreases (consequently, $G_{\it eff}$ increases) whilst the scale factor
increases (where always $\ddot{a}<0$). The solutions associated to $\lambda_-$
and $\lambda_+$ are known as slow and fast solutions, respectively~\cite{Faraoni.book}.
Under the well-known duality transformation associated to the spatially flat FLRW
metric for the BD cosmology in vacuum~\cite{dual1,dual2},
it has been shown that the slow and fast solutions are interchanged \cite{Faraoni.book}.
Moreover, we should note that for both the cases (i) and (ii), when the BD coupling parameter
is restricted to $-3/2<\omega<-4/3$,  we can easily
show that $\phi(t)$ always decreases with the cosmic time, while the
scale factor always accelerates for the case (i) and decelerates for the case (ii).
These non-deformed solutions will be compared with the corresponding NC cases in
 this paper. Furthermore, it should be emphasized that
 for all of these cases, there is a big bang singularity as $t\rightarrow t_i$.

By assuming $t>t_i$, when the constant $C$ can take negative
values, it is worthwhile to look at the above solutions from another perspective.
Again, let us assume that $\omega$ is restricted to $-4/3<\omega<0$.
 In this case, the solutions associated to $\lambda_-$ correspond to a pre-big bang
 inflationary scenario where the scale factor $a(t)$ accelerates.
 In this case, $\phi(t)$ decreases as cosmic time progresses.
 However, we must obtain an expanding universe after the big bang and also
 we should respect to the constraints on the evolution of the $G_{\it eff}$.
 Consequently, the solutions with such properties are merely provided by considering the solutions case (ii)
 described above. Namely, to get post-big bang solutions, we need
 a branch changing, i.e., $\lambda_-\rightarrow \lambda_+$, as well as $C<0\rightarrow C>0$.
 These pre and post-big bang solutions include the solutions obtained in \cite{C98} and correspond to low energy limit
 of some string theories, namely, the particular value of the BD coupling parameter $\omega=-1$.
\subsection{General deformed case with $\zeta(a)=a^n$}
Henceforward, let us proceed our investigation with only power-law
functions of the scale factor in the deformed Poisson bracket, namely, $\zeta(a)=a^n$ and
investigate the corresponding time behaviors for the physical quantities.
First, we will obtain general equations of motion and
subsequently, we will show how specific values of $n$ produce interesting dynamics.

%\subsection{ $\zeta(a)=a^n$}
%\bl{Before proceeding the calculations of this section, let us discuss regarding the
%deformed Poisson bracket associated to this general power-law case.
%From (\ref{NC-Poisson}), we obtain
%\begin{eqnarray}\label{NC-Poisson-2}
 %\{P_a,P_\phi\}=\theta a^n \phi,
%\end{eqnarray}
%and other Poisson brackets given by (\ref{NC-Poisson}) remain unchanged.
%Substituting the relation for the scale factor from (\ref{H-a}) into (\ref{NC-Poisson-2}), we obtain
%\begin{eqnarray}\label{NC-Poisson-3}
 %\{P_a,P_\phi\}=\theta a^n\left(\frac{a}{a_0}\right)^{\frac{1}{\chi}},
%\end{eqnarray}
%which indicates that the BD coupling parameter plays very important role in NC portion of our model.
%In the particular case where $|\omega|$ goes to
%infinity\footnote{\bl{regarding this case in BD I should add a comment for trace free ordinary EMT}}, then $1/\chi$ vanishes and therefore we get
%\begin{eqnarray}\label{NC-Poisson-4}
% \{P_a,P_\phi\}=\theta a^n.
%\end{eqnarray}
%Consequently, when $n=0$, we get $\{P_a,P_\phi\}=\theta$, which has been completely investigated in \cite{RFK11}.
%However, in this paper, we would study the case $n=0$ in general rather than only for large values of the BD coupling parameter.
%Moreover, relation indicateas }

By substituting $\zeta=a^n$ into (\ref{phidot-2}) and using (\ref{H-a}), it is
easy to show that
%\footnote{Hereafter, we will use index $n$  in $f_n(\phi)$ for
%denoting the same values employed in the power-law function of the scale factor.}
\begin{eqnarray}\label{phidot-gen}
\dot{\phi}=-\frac{1}{a_i^3 \phi^{3\chi}}\left\{b+\frac{a_i^{n+1}\theta(2\chi+1)\phi^{[(n+1)\chi+2]}}
{6\xi^2[(n+1)\chi+2]}\right\}
\equiv f_n(\phi),
 \end{eqnarray}
where we have assumed general values of the BD coupling parameter where $(n+1)\chi\neq-2$.
 Similarly, by substituting $\zeta(a)=a^n$ and $\dot{\phi}$
 from (\ref{phidot-gen}) into (\ref{a2dot-2}), we obtain
 \begin{eqnarray}\label{a2dot-case1}
\frac{\ddot{a}}{a}=\frac{3\chi-\omega}{3}
\left[1+\frac{\theta a_i^{n-2} \phi^{[(n+1)\chi+2]}}{6\xi^2 f_n(\phi)}\right]\left[\frac{f_n(\phi)}{\phi}\right]^2.
\end{eqnarray}
Moreover, the energy density and the pressure associated to the BD scalar field are given by
\begin{eqnarray}\label{ro-3}
\rho_\phi\!&=&\!\frac{\omega-6\chi}{16\pi\phi}\left[f_n(\phi)\right]^2,
\\\nonumber
\\
p_\phi\!&=&\!\frac{1}{8\pi}\left[\frac{(\omega-2\chi)\phi}{2}
+\frac{a_i^{n-2}\theta(\omega-3\chi)}{9\xi^2}\frac{\phi^{[(n-2)\chi+3]}}{f_n(\phi)}\right] \left[\frac{f_n(\phi)}{\phi}\right]^2,\label{pr-3}
\end{eqnarray}
where we have used (\ref{ro-2})-(\ref{H-a}) and (\ref{phidot-gen}).

To obtain an accelerating universe the following condition must be satisfied
\begin{eqnarray}\label{acc-con}
\frac{3\chi-\omega}{3}
\left[1+\frac{\theta a_i^{n-2} \phi^{[(n+1)\chi+2]}}{6\xi^2 f_n(\phi)}\right]>0,
\end{eqnarray}
which requires determining the BD scalar field in terms of the cosmic
time, which, in turn, is obtained from solving differential equation (\ref{phidot-gen}).
Moreover, we should note that to avoid obtaining ghost models, we must
check the values taken by the kinetic energy density according to relation (\ref{ro-3}).
By considering the Einstein representation of the standard BD models, for an FLRW line-element, as long as the
BD coupling parameter is restricted to $\omega\geq-3/2$, the kinetic energy density takes positive values~\cite{Lev95}.
We will show, for the X-branch solutions, how the above constraints are satisfied for a short time and then
it can exit from the acceleration phase and then turns to a decelerating one.

In this paragraph, let us discuss the simplest case of deformation in the phase space.
 More concretely, by assuming $\zeta=1$, (\ref{NC-Poisson}) yields
  \begin{eqnarray}\label{GRG-1}
 \{P_a,P_\phi\}=\theta\phi.
\end{eqnarray}
With a quick glance at relations (\ref{NC-Poisson}) and (\ref{H-a}), it can be easily indicated that
BD coupling parameter plays a very important role in our herein deformed model.
For instance, for this simplest case $n=0$, when $|\omega|$ goes to infinity\footnote{We should note that in the
particular case of our NC model where $\omega\rightarrow\infty$,
then $\phi$ does not take constant values and therefore relation (\ref{NC-Poisson-4}) does make sense.},
then $1/\chi$ vanishes and therefore we get
\begin{eqnarray}\label{NC-Poisson-4}
 \{P_a,P_\phi\}=\theta.
\end{eqnarray}
This very particular case has been investigated in \cite{RFK11}.
In that paper, by proposing a special kind of
 deformation in the Poisson bracket associated to the momenta of
  the BD scalar field and the FLRW scale factor, in the particular case
  where $|\omega|\rightarrow\infty$, interesting
 results have been obtained. For instance, it has been shown
 that it is possible to overcome the horizon and the graceful exit
 problems for an accelerating epoch associated to the early times of the universe.
 However, it has been neither discussed quantitatively
 nor the general values of the BD coupling parameter were employed to solve the horizon problem.
 Further investigations on this simplest case will not be of interest in this paper.
 In this work, not only we will analyze more
 general solutions in which the BD coupling parameter as well as (integer) $n$ take arbitrary values
  but also we will investigate the horizon problem quantitatively.
  Furthermore, the cosmological phase
  space analysis as well as the discussions on the EF results will be in our focus.

 % Let us solve the equations of motion for both the commutative and NC cases.
%  In the commutative case where $\theta=0$, we can easily solve equation (\ref{}) and obtain an exact general solution as
 % \begin{eqnarray}\label{com-phi-a}
 % \phi(t)&=& c_2 \left[(3 \chi+1) t-c_1\right]{}^{\frac{1}{3 \chi+1}},\\\nonumber
 % \\
 % a(t)&=&a_0c_2^\chi \left[(3 \chi+1) t-c_1\right]{}^{\frac{\chi}{3 \chi+1}},
  % \end{eqnarray}
% where $c_1$ and $c_2$ are the integration constants. In the
%particular case where $\omega\rightarrow\infty$, then the BD scalar field
%takes a constant value, namely, we get $\phi=c_2$ , and consequently,
%the scale factor becomes $a(t) =$. Tt
%has been shown that when the trace of the ordinary mat-
%ter vanishes in the standard BD theory, the correspond-
%ing GR solutions may not recovered [].

\subsubsection{Particular exact solutions of the deformed case with $\zeta(a)=a^n$}
First, let us discuss concerning the exact analytic solutions which can be
obtained from (\ref{phidot-gen}) for the particular cases and subsequently
analyze the solutions for the more general cases.
\begin{itemize}
  \item {\bf Class I: $\chi\neq1/(2-n)$}\\

  In this case, when $b\neq0$, solving differential equation (\ref{phidot-gen}) leads to
 the hypergeometric functions, which is complicated to deal with as exact
  analytic solutions. However, we will discuss this case when we will employ a numerical analysis.
 In the particular case where $b=0$, it is straightforward to show that
\begin{eqnarray}\label{p-phi-power1}
t-t_i=\frac{6a_i^{2-n}[(n+1)\chi+2]\xi^2}{[(n-2)\chi+1](2\chi+1)\theta}\phi^{[(2-n)\chi-1]},
 \end{eqnarray}
 where $t_i$ is an integration constant.

  \item {\bf Class II: $\chi=1/(2-n)$ where $n\neq2$ }\\\\
    In this case, from relations (\ref{zi}) and (\ref{H-a}),
    we obtain constant values for $\omega$ and $\xi$ as
    \begin{eqnarray}\label{cons-omega}
\omega=\frac{6(3-n)}{(n-2)^2}, \hspace{10mm} \xi=\frac{(4-n)\lambda}{n-2}.
 \end{eqnarray}
    Similar to the previous class, we obtain different solutions correspond
     to whether $b\neq0$ or $b=0$. Namely, when $b\neq0$, we obtain an exact solution as
 \begin{eqnarray}
t-t_i&=&\left[\frac{6a_i^{2-n}(n-4)}{(2-n)\theta}\right]
\,{\rm ln}\left[1+\frac{a_i^{n+1}(n-2)^2\theta}{6b(n^2-9n+20)}\phi^{\frac{n-5}{n-2}}\right],\label{phi-log-1}
 \end{eqnarray}
and for $b=0$, we get
 \begin{eqnarray}\label{phi-log-2}
t-t_i=-\left[\frac{6a_i^{2-n}(n-4)(n-5)}{(2-n)^2\theta}\right]{\rm ln}\phi,
 \end{eqnarray}
\end{itemize}
 It is easy to rewrite the relations (\ref{p-phi-power1}), (\ref{phi-log-1}) and (\ref{phi-log-2}) such that the
 BD scalar field is stated in terms of the cosmic time.
Consequently, for these set of exact solutions, from (\ref{H-a}), we can obtain exact solutions
for the scale factor versus the cosmic time.

%In what follows, we will proceed to analyse the solutions associated to the specific values of $n$.

\subsubsection{General numerical results of the deformed case with $\zeta(a)=a^n$}
  Unfortunately, for the general NC case, it is almost impossible to obtain analytic exact solutions.
Therefore, let us discuss this case by employing a numerical analysis.
For the sake of completeness, we will compare our NC results
with those obtained from standard exact solutions already reported.
 In what follows, let us analyze the time behavior of the quantities arising from the
   numerical analysis, associated to the solutions for both the commutative and NC case.
 Various numerical studies have produced the following interesting results.

\begin{description}
 \item[{\bf Lower solutions ($\lambda=\lambda_-=-1$):}] We have already demonstrated that for the non-deformed
 case, with the values of the BD coupling parameter from interval $-3/2<\omega<-4/3$ and
 assuming $C>0$, we do obtain accelerating scale factor.
 Similarly, by applying our herein NC model, by employing numerical codes, we could find more larger
 intervals of $\omega$ (with respect to the standard case)
 where the scale factor always accelerates\rlap.\footnote{As the D-branch
 solutions will not be the scope of the investigation of the present
 work, let us forbear from plotting the time behavior of the physical quantities.
 In \cite{RM14}, by introducing another NC model, the D-branch solutions have been completely investigated.}
  However, a mere accelerating scale factor cannot be considered as a sufficient condition for a successful
 scenario for an inflationary epoch of the early universe.

 Let us be more precise.
 In this case, notwithstanding applying several numerical endeavors, we
 could not find any condition to solve the graceful exit problem.
 Indeed, this is the most important problem with the D-branch
 solution in BD theory (even with varying $\omega$ \cite{Lev95-2}) as well
 as in the string theory \cite{BV94}.
 Obviously, as will be shown in the next section, the requirement to
  either satisfying the sufficient condition or fully
 overcome the graceful exit problem can be directly related to
 the time behavior of either $a \phi^{1/2}$ (associated to the JF) appearing in
 the sufficient condition (the most strongest
 condition which is required to solve the horizon problem in the standard
 cosmology) or the scale factor associated to the EF.
 More concretely, in \cite{Lev95-2}, it has been
 emphasized that (i) the D-branch solution suffers from the graceful exit
 problem in the kinetic inflationary scenario in the BD theory \cite{Lev95-2} as well
 as in the string theory \cite{BV94}, namely, the corresponding
  inflation cannot exit from the accelerating phase
 successfully, to enter an expanding universe (with $\ddot{a}<0$);
 (ii) for the X-branch solution, there may be a
 possibility to exit it and subsequently to enter into an
 expanding phase, but, this solution suffers from the flatness problem.
 Moreover, it has been mentioned that
 the problems mentioned in (i) and (ii) are due to the lack of
 ability of those models to obtain an accelerating scale
 factor in the corresponding EF \cite{Lev95-2}.
 Nevertheless, in \cite{RM14}, by employing
 another noncommutativity for the BD theory, and focusing
 on the D-branch solutions, we have obtained a
 successful kinetic inflation without encountering the
 problem mentioned in (i).

 In the present work, for our chosen NC framework, it seems that we
 cannot overcome the problems associated to the D-branch solutions.
Instead, in what follows, let us focus on the X-branch solutions.
We will show that the herein NC model can provide
conditions to obtain X-branch solutions which can be considered
as a successful kinetic inflationary scenario.

  \item[{\bf Upper solutions ($\lambda=\lambda_+=+1$):}]
   %In what follows, let us just
%summarize the interesting results associated to the
%NC case, which are completely different from cor-
%responding ones obtained in the standard case.
  Before reporting our numerical results
  for this case, we should mention a few important points.
  (i) In order to analyze and depict
  the time behavior of the quantities, except the NC parameter, we have used
  the same ICs for both the commutative
  and NC cases to solve equations (\ref{H2})-(\ref{phi2dot}) by taking constant
  values of the BD coupling parameter and $\zeta=a^n$.
  Obviously, we just need the initial values for $\phi$, $a_i$ and $\dot{\phi}$ [which is equivalent to the integration
  constant $b$ in equation (\ref{phidot-gen})].
  (ii) We have assumed that $C > 0$, and $-4/3<\omega<0$ in relations (\ref{OT-phi1}) and (\ref{OT-a1})
for the standard solutions and $-3/2<\omega<0$ for the NC cases.
(iii) We have investigated this case by taking very small positive values for the NC parameter $\theta$.
 (iv) $n$ is restricted as $n\leq 0$ such that $|n|$ should not take very large values.
 (v) We found that for every set of parameters \{$\omega$, $n$, $\theta$\}
 (which should satisfy the conditions (ii), (iii), (iv)), we can always choose
appropriate set of reasonable initial conditions \{$a_i$, $\phi(0)$ $\dot{\phi}(0)$\},
which can produce the following interesting consequences.
\begin{itemize}
  \item
  For the NC case, we have shown that the BD scalar field has a slightly
   different time behavior in comparison to the standard BD model at early times. Let us be more precise.
 We found that for both cases, $\phi$ contracts such that for
 all times $\dot{\phi}<0$ and $\ddot{\phi}>0$, such that for the commutative
 case $\ddot{\phi}$ always decreases when the cosmic time grows, see Figs. \ref{phi-up}.
  However, with taking any values of the BD coupling
  parameter such that $\omega>-3/2$, taking the same ICs as for the commutative case, and employing very small positive values of the
  NC parameter, we have shown that the BD scalar field always contracts such that for
  the early times $\ddot{\phi}$ increases. Whilst, after a short time, it turns to
decrease.
  Moreover, for large values of the cosmic time, $\ddot{\phi}$ asymptotically vanishes; for instance, see Fig. \ref{phi-up}.

  \begin{figure}
\centering\includegraphics[width=2.5in]{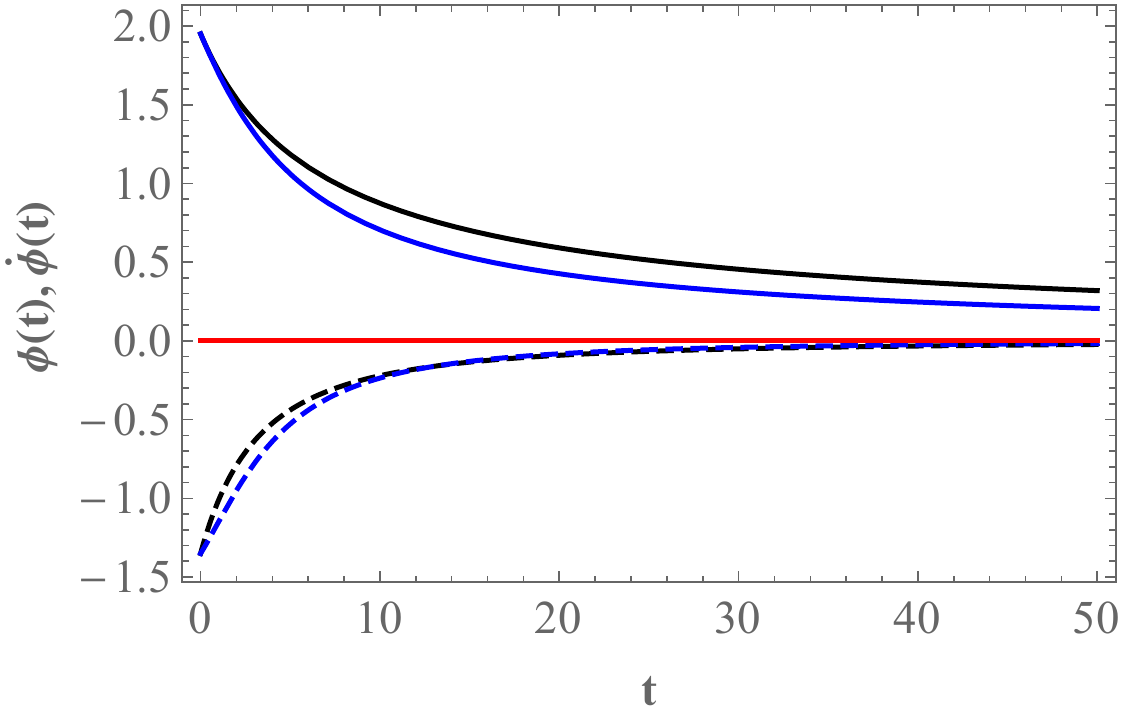}
\hspace{5mm}
\centering\includegraphics[width=2.5in]{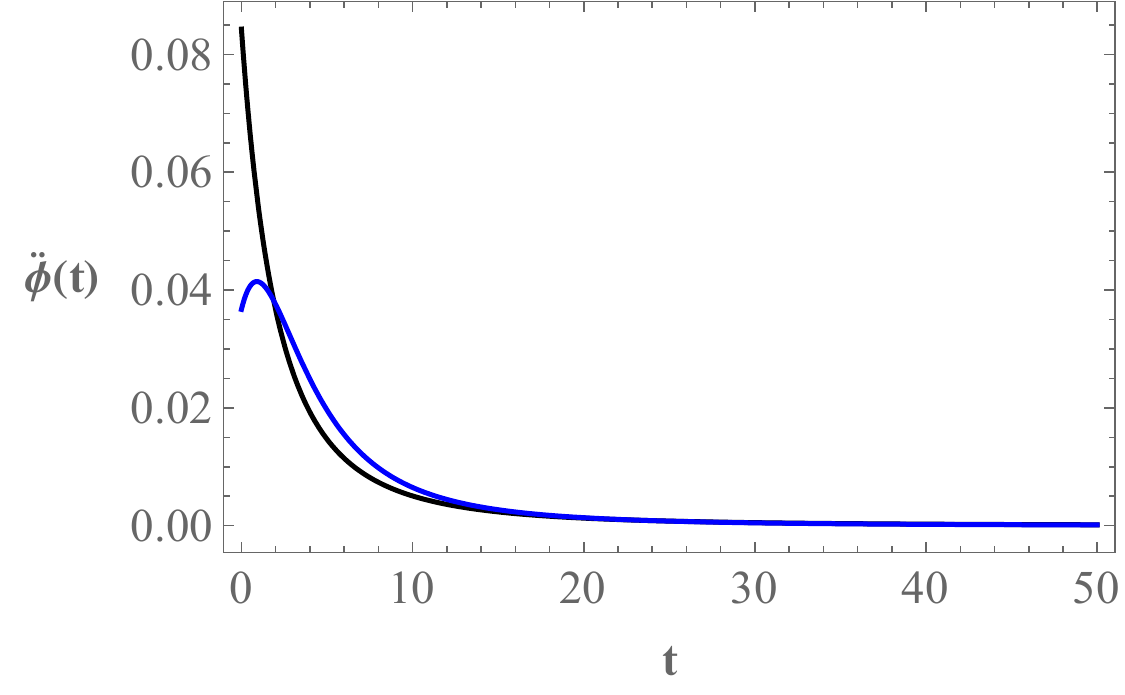}
\caption{The time behavior of $\phi(t)$ (solid curves in left panel),
$\dot{\phi}(t)$ (dashed curves) and $\ddot{\phi}(t)$ (right panel) for the commutative case
(black curves) and NC case (blue curves) for the upper solution ($\lambda=+1$).
We have set $\omega=-1.12$, $a_i=0.01$, $\phi(0)=1.95$, $\dot{\phi}(0)=-0.27$,
$n=0$ and $\theta=1\times10^{-5}$ (for the NC case). The red line is for $\phi(t)=0$.
 For greater visibility, we re-scaled the plots of $\dot{\phi}$. We have used the Planck units. }
\label{phi-up}
\end{figure}

 \item
  For the commutative case, the scale factor of the universe
  always decelerates. Namely, for all the times, the universe expands with
  $\ddot{a}<0$. Let us see that how the effects of the
  noncommutativity change the behavior of the scale factor.
  For the same above mentioned ICs, we see that the scale
  factor associated to the NC case accelerates for very early times, namely, $\ddot{a}>0$.
  Subsequently, after a short time, it turns to decelerate.
  Finally, for very large values of the cosmic time, our numerical endeavors
  show that we get zero acceleration for the scale factor of the universe. Moreover, for the deformed case,
  we see that the big bang singularity is removed; see, for instance, Figs. \ref{a-up}.
 \begin{figure}
\centering\includegraphics[width=2.5in]{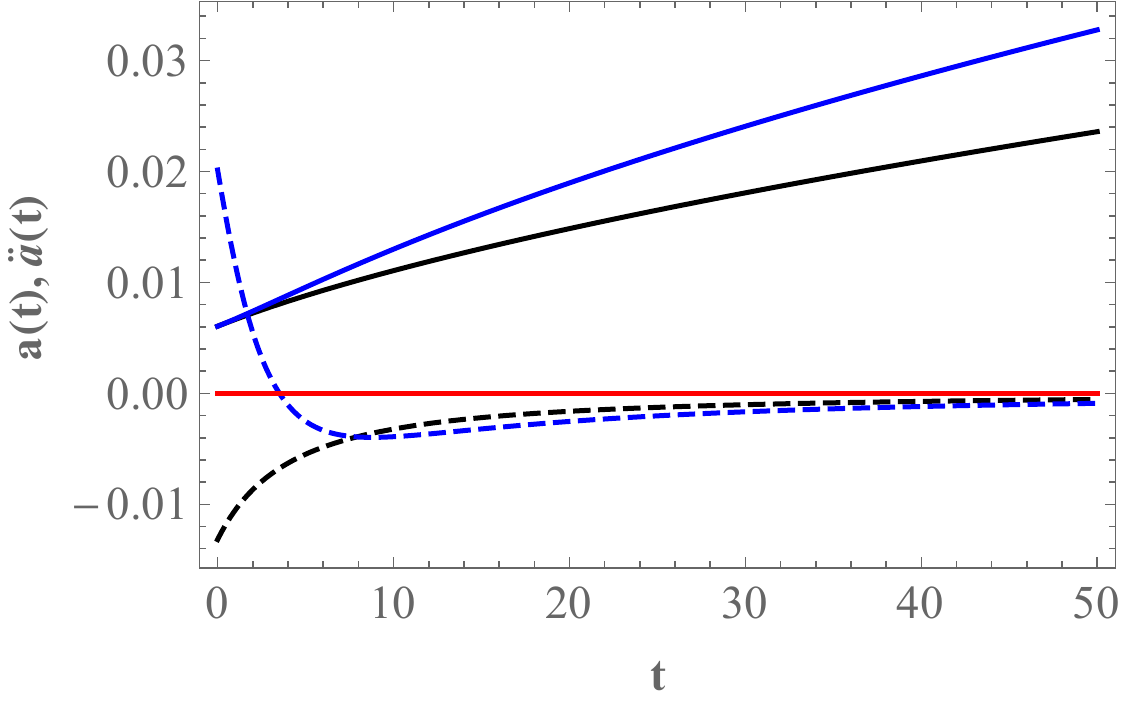}
\hspace{5mm}
\centering\includegraphics[width=2.5in]{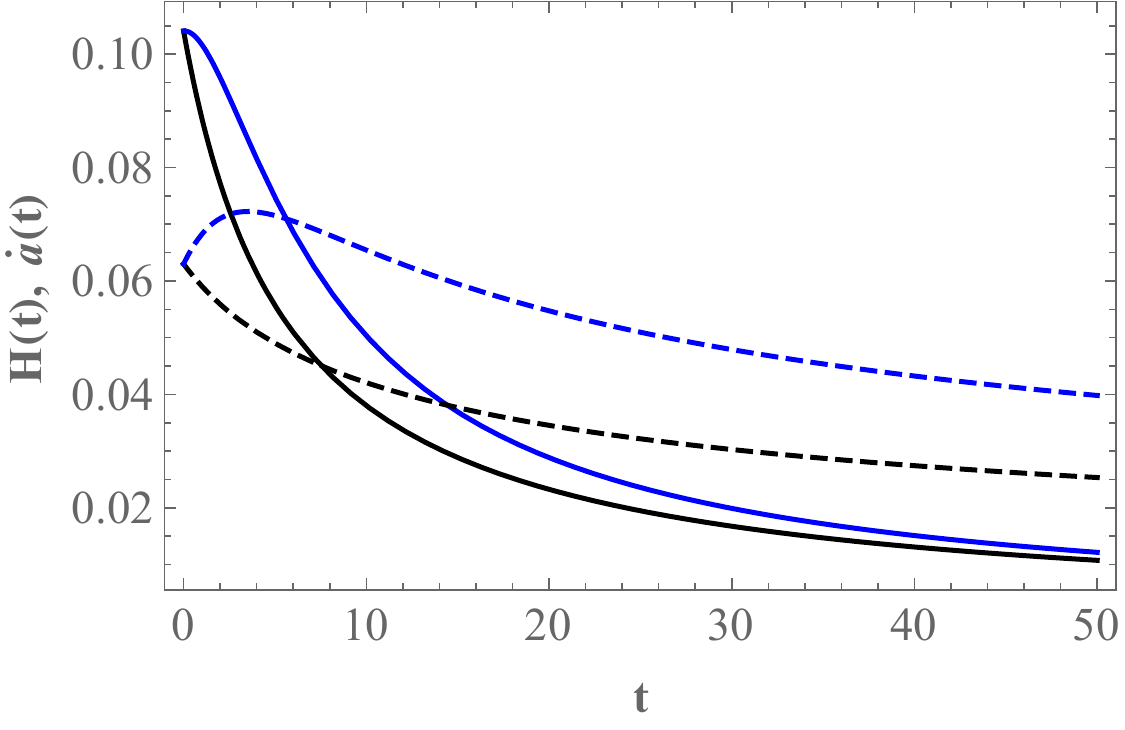}
%\centering\includegraphics[width=3in]{a2dot-up}
\caption{ Left panel: the time behavior of the scale factor (solid curves) and its
second time derivative (dashed curves). Right panel: the time behavior of the Hubble parameter
(solid curves) and $\dot{a}(t)$ (dashed curves). The black and blue curves are associated to the commutative case
and NC case, respectively.
We have set $\lambda=+1$, $\omega=-1.12$, $a_i=0.01$, $\phi(0)=1.95$, $\dot{\phi}(0)=-0.27$,
$n=0$ and $\theta=1\times10^{-5}$ (for the NC case). The red line is for $a(t)=0$.
 For greater visibility, we re-scaled the plots of $\dot{a}$ (not $H(t)$) and $\ddot{a}$.
 We have used the Planck units. }
\label{a-up}
\end{figure}
 \item
 For the commutative case, the numerical results show that both the
energy density and the pressure associated to the BD scalar field decrease
with the cosmic time. Moreover, both of them get positive values for all the
times of their evolution.
 For the NC case, like the commutative case, the energy density always
 gets positive values.
 However, its time behavior is different from the commutative case. More concretely, contrary to
 the commutative case, for which $|\dot{\rho}_\phi|$ always
 decreases with cosmic time, for the NC case, it
 increases such that after getting
 a maximum value turns to decrease. Moreover, for large values of the cosmic time it tends to zero.
However, the NC pressure always takes negative values
such that it increases with the cosmic time, and similar to NC energy density, it asymptotically approaches to zero.
As an example, in Fig. \ref{ro-p-up}, the energy density and pressure
associated to the BD scalar field have been plotted in terms of the cosmic time.
 Moreover, our numerical endeavors have shown that the equation of
state (EoS) parameter $W_\phi\equiv p_\phi/\rho_\phi$ takes always a constant positive value for the commutative case.
Whilst, for the NC case, it evolves during the inflationary epoch (such that $W_\phi (t)<-1/3$), and
approaches to its corresponding commutative value for late times, see, for instance Fig. \ref{ro-p-up}.
Should it be emphasized that, throughout this paper, among many sets of reasonable ICs which could yield
inflation for the X-branch solutions with different ranges of the EoS
parameter, we have restricted our attention to those which do produce successful
inflationary phase provided that the EoS parameter is restricted to the viable range $(-1,1)$.

\end{itemize}
 \begin{figure}
\centering\includegraphics[width=2.5in]{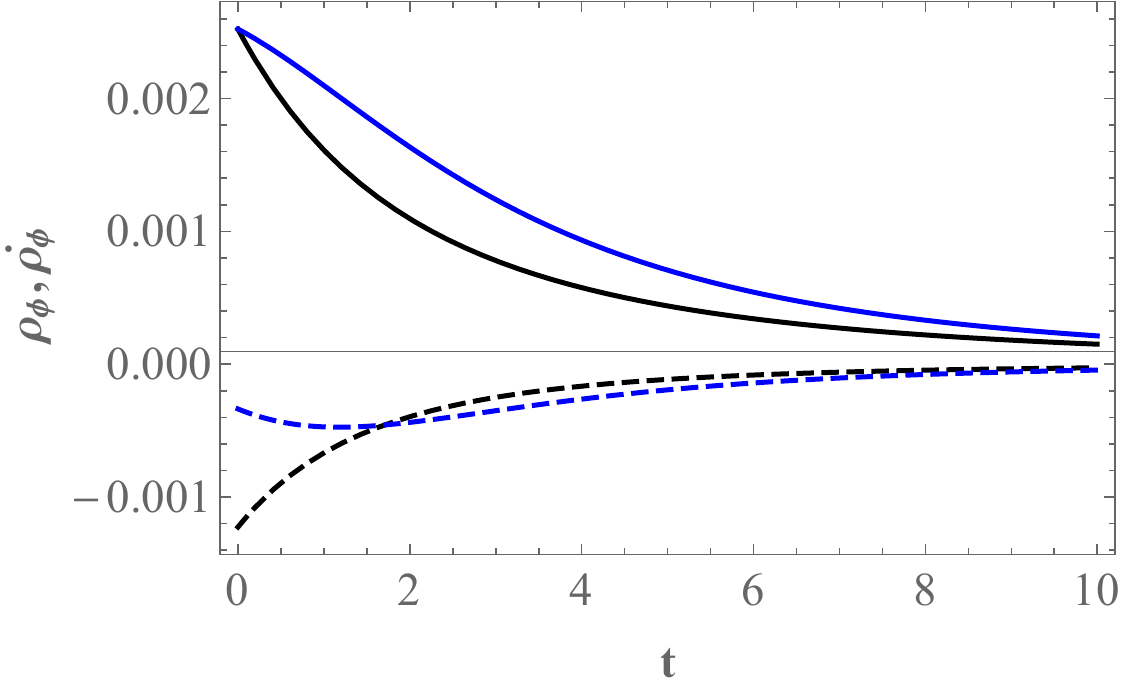}
\hspace{5mm}
\centering\includegraphics[width=2.5in]{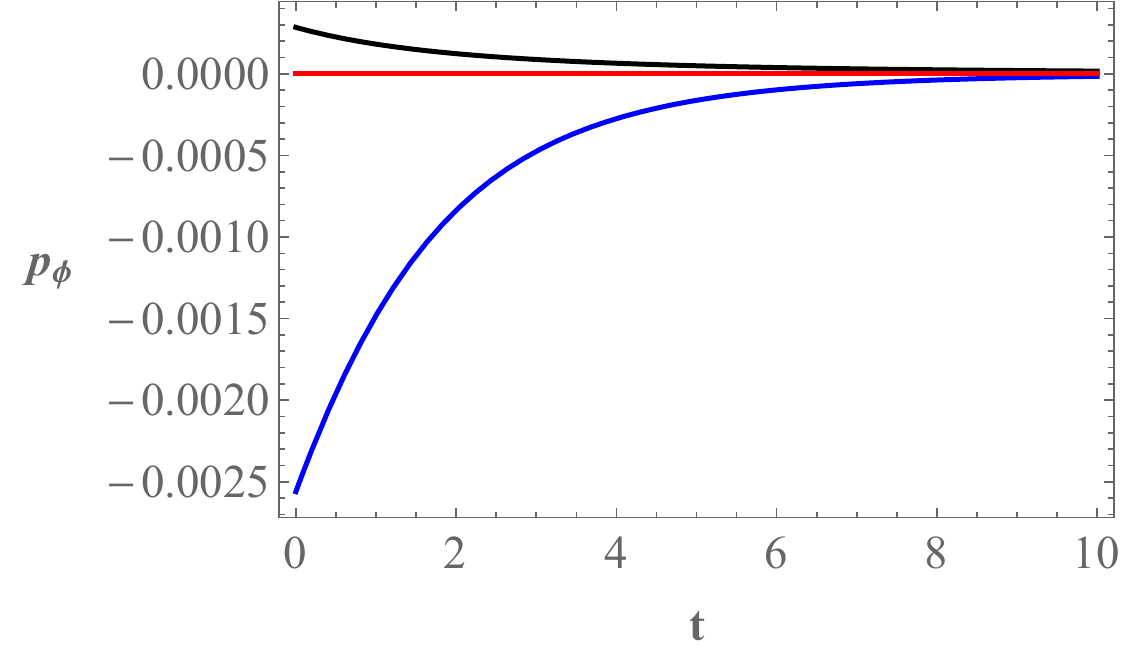}
\centering\includegraphics[width=2.5in]{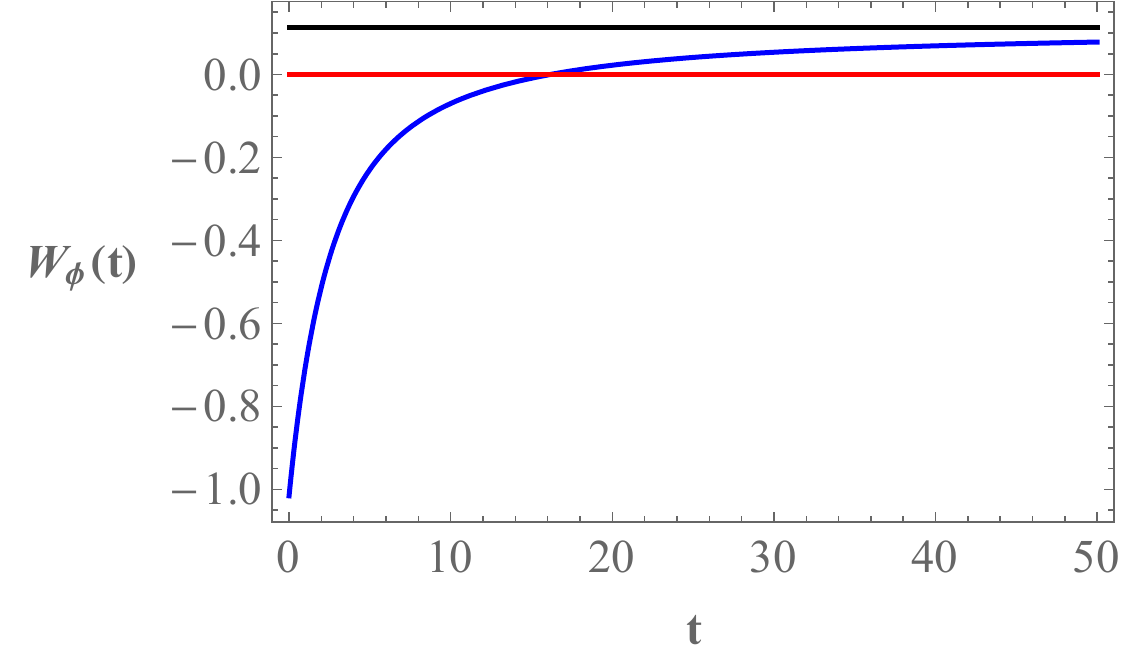}
\caption{Upper left panel: the time behavior of $\rho_\phi$ (solid curves) and $\dot{\rho}_\phi$ (dashed curves).
Upper right panel: the time behavior of pressure. Lower panel: the
time behavior of the EoS parameter.
The blue and black curves are associated to the NC and
commutative cases, respectively, and red lines are for $p_\phi=0=W_\phi$.
We have set $\lambda=+1$, $\omega=-1.12$, $a_i=0.01$, $\phi(0)=1.95$, $\dot{\phi}(0)=-0.27$,
$n=0$ and $\theta=1\times10^{-5}$ (for the NC case).
 We have used the Planck units.}
\label{ro-p-up}
\end{figure}

\item
Let us now describe how for different values of $n$, the NC parameter and the BD
coupling parameter affect the time behavior of the physical quantities.
 In order to show that our general results listed above
 can also be obtained with extended ranges of the suitable ICs, we would like to plot
 the following figures with a suitable set of different values of the ICs.
Our numerical efforts indicate that the different values of either the positive NC
parameter or negative BD coupling parameter or $n$ do not affect
on the resulted background features explained above
(for instance, getting accelerating scale factor at early times and then decelerating one).
However, by taking different values of these parameters, the time
interval of acceleration as well as the corresponding amounts of
the physical quantities effectively are changed. For instance, we have shown that during inflation the
smaller $\theta$ (or $|\omega|$ or $|n|$), the smaller the corresponding time interval of inflation
and slope of the scale factor.
In Figs. \ref{adot-dif-par-up}
and \ref{adot-dif-n-up}, the time behavior of $\dot{a}$ has been
shown for different values of $\omega$, $\theta$ and $n$.
Our numerical results have shown that for the X-branch, similar to the case $n=0$, we
can get a viable inflation with $-1<W_\phi<1$, see for instance, Fig. \ref{adot-dif-n-up}.
\begin{figure}
\centering\includegraphics[width=2.5in]{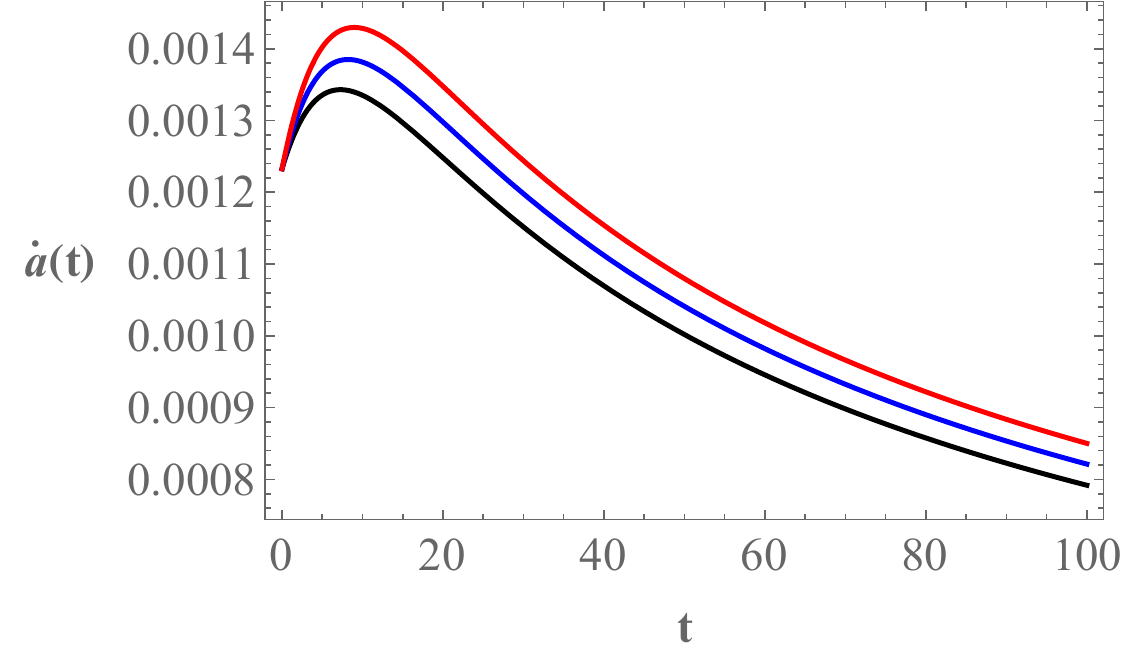}
\hspace{5mm}
\centering\includegraphics[width=2.5in]{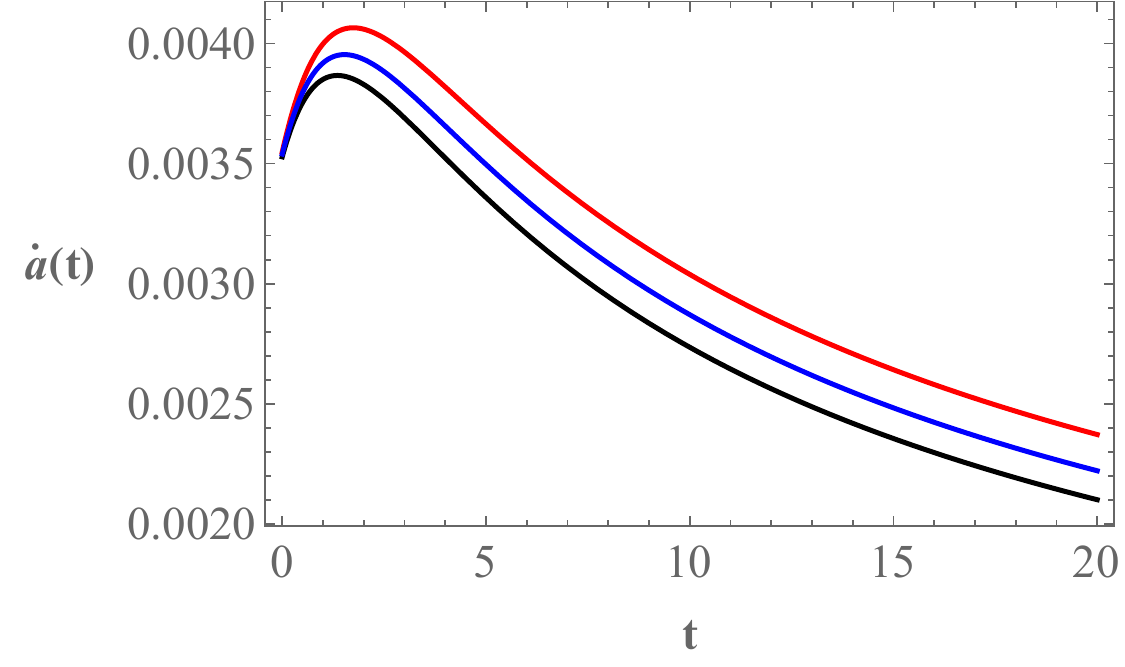}
%\centering\includegraphics[width=3.2in]{adot-dif-n}
\caption{ The time behavior of $\dot{a}$ for the deformed case and
the upper solution ($\lambda=+1$) for different values of
(i) the NC parameter (left panel): $\theta=8\times10^{-5}$ (black curve), $\theta=9\times10^{-5}$ (blue curve) and
$\theta=1\times10^{-4}$ (red curve) where we
 have set $\omega=-1.1$, $n=0$, $a_i=0.05$, $\phi(0)=2$ and $\dot{\phi}(0)=-0.11$;
  (ii) the BD coupling parameter (right panel):
$\omega=-0.8$ (black curve), $\omega=-0.9$ (blue curve) and $\omega=-1$ (red curve),
where $\theta=1\times10^{-4}$, $n=0$, $a_i=0.05$, $\phi(0)=3.8$ and
$\dot{\phi}(0)=-0.98$.We have used the Planck units.
\label{adot-dif-par-up}}
\end{figure}

\begin{figure}
\centering\includegraphics[width=2.5in]{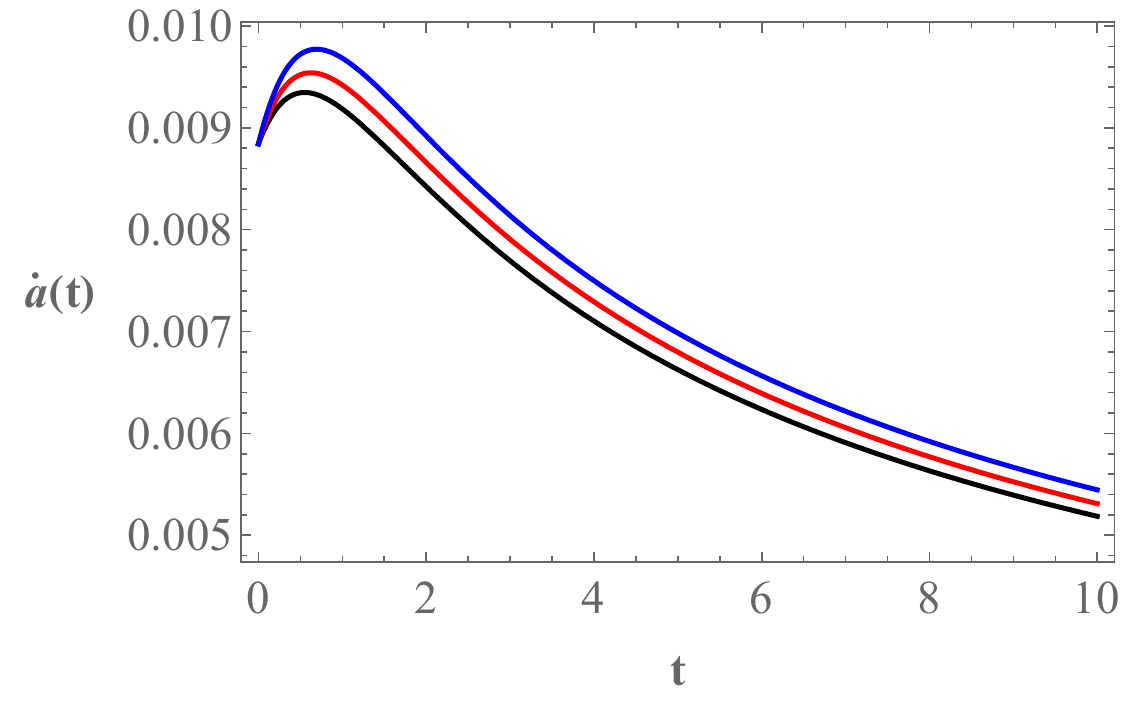}
\hspace{5mm}
\centering\includegraphics[width=2.5in]{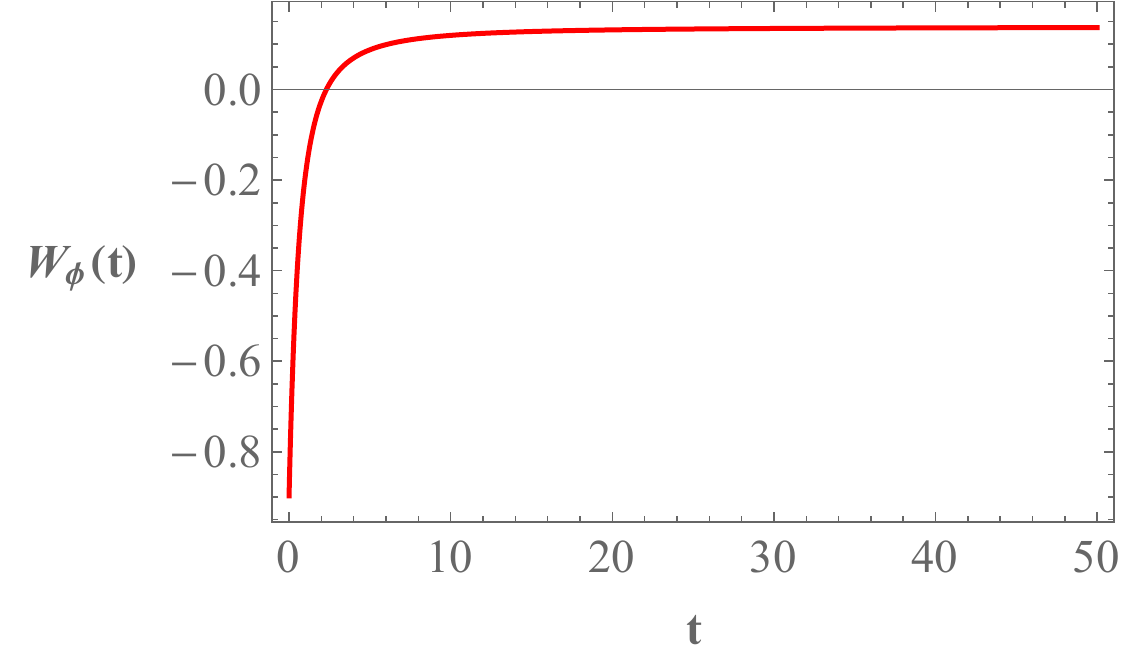}
%\centering\includegraphics[width=3.2in]{adot-dif-n}
\caption{ The time behavior of $\dot{a}$ (left panel)
and $W_\phi$ (right panel, for $n=-1$) for different values of $n$.
 We have set $n=-0.97$ (black curve),
$n=-1$ (red curve), $n=-1.03$ (blue curve) and $\lambda=+1$, $\omega=-1.05$, $a_i=0.0515$,
$\phi(0)=2.65$, $\dot{\phi}(0)=-1.25$, $\theta=1\times10^{-5}$.
The time behavior of $W_\phi$ for other values of $n$ are almost
similar to that of $n=-1$. We have used the Planck units.
 \label{adot-dif-n-up}}
\end{figure}

\end{description}
%%%%%%%%%%%%%%%%%%%%%%%%%%%%%%%%%%%%%%%%%%%%%included JMarto%%%%%%%%%%%%%%%%%%%%%%%%%%%%%%%%%%%%%%%%%%%%%%%%%%%%%%%%%%%%%%%%%%%%%%%

\begin{figure}
\centering\includegraphics[width=2.5in]{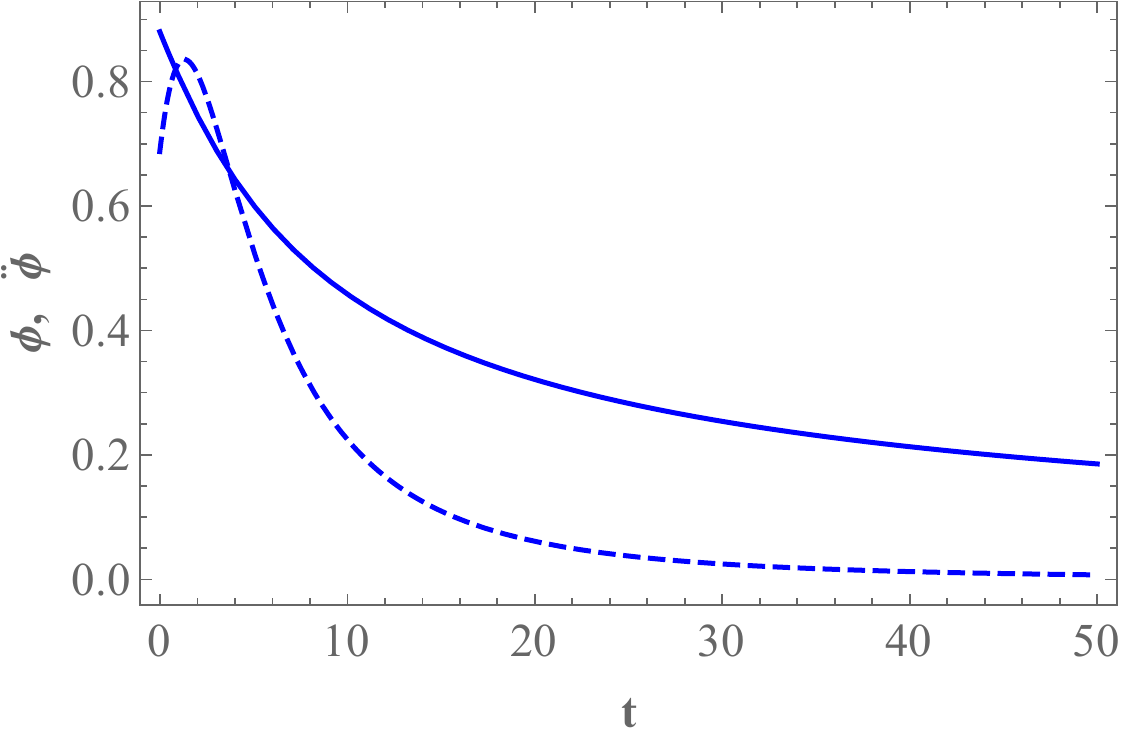}
\hspace{5mm}
\centering\includegraphics[width=2.5in]{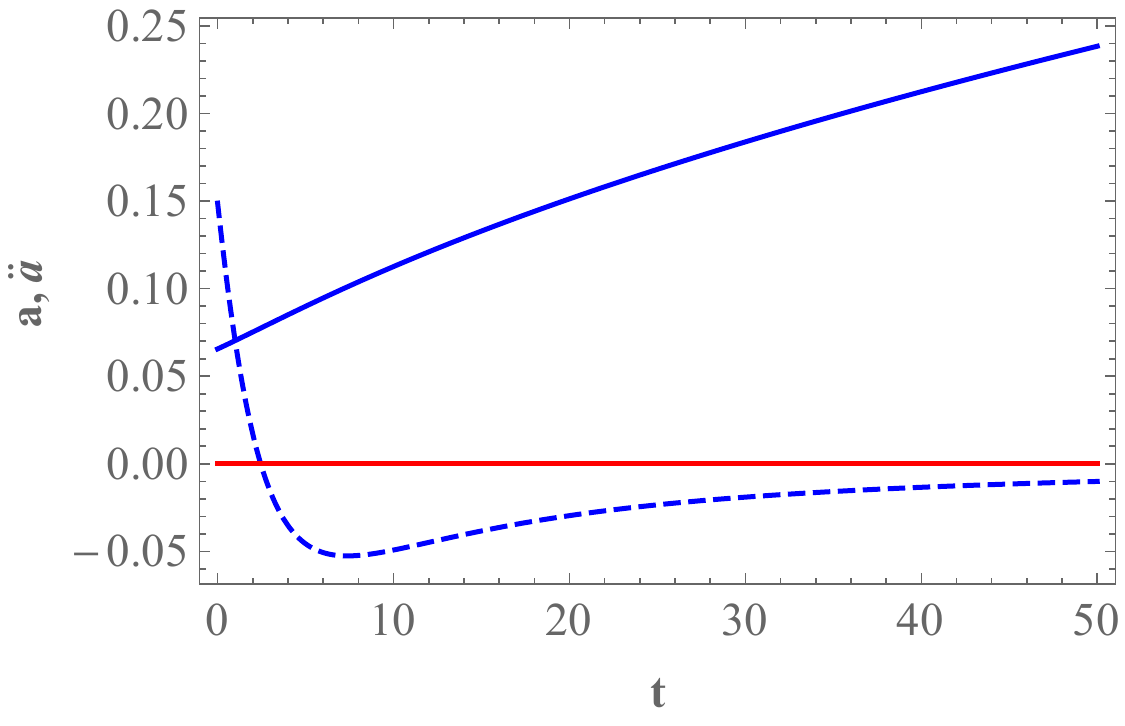}
\centering\includegraphics[width=2.5in]{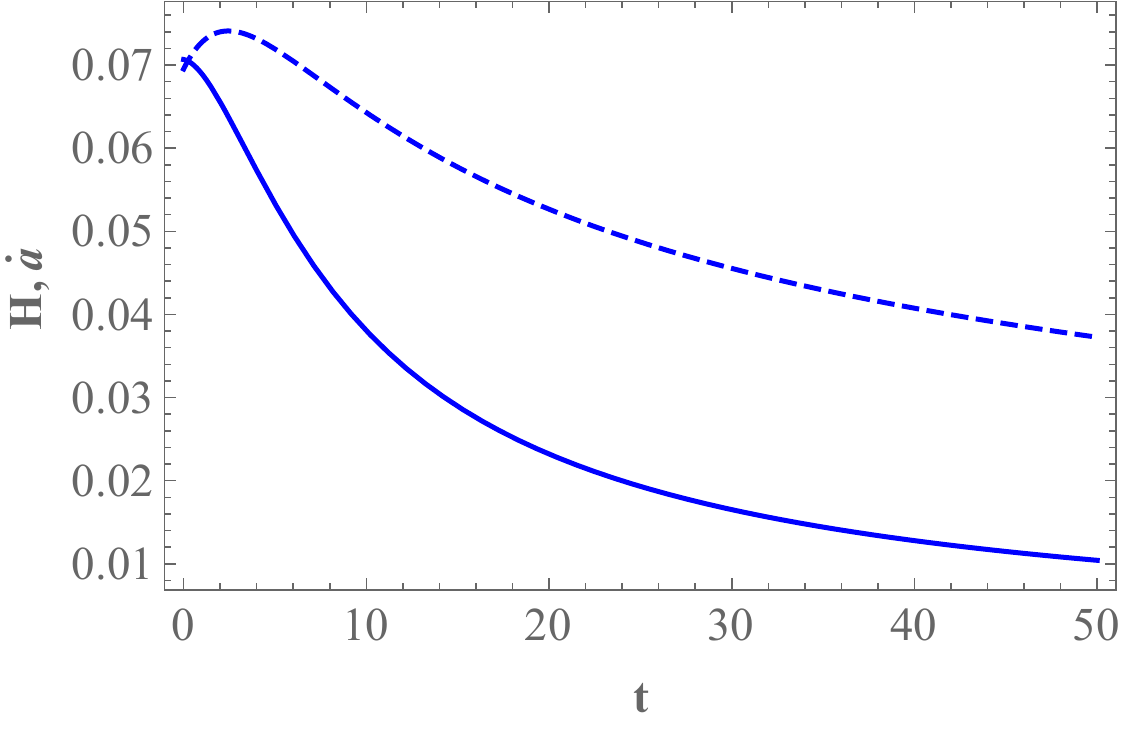}
\hspace{5mm}
\centering\includegraphics[width=2.5in]{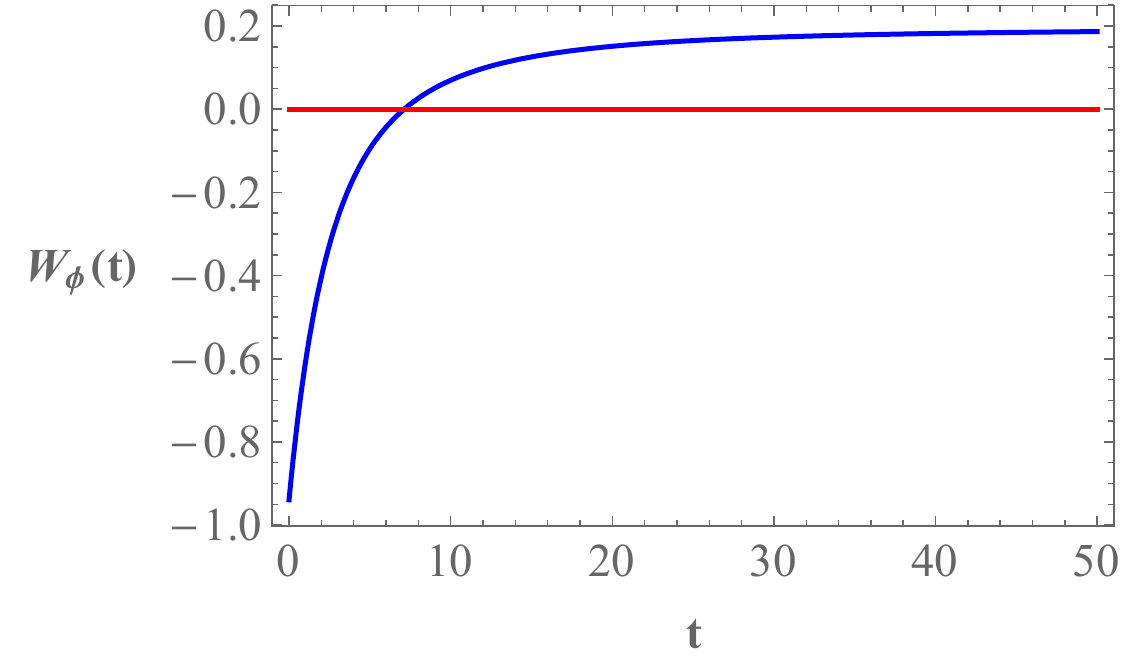}
%\centering\includegraphics[width=3in]{a2dot-up}\[Theta]0 := 0.00002
\caption{ The solid blue curves are associated to the time behavior
of $\phi(t)$, $a(t)$, $H(t)$ and $W_\phi(t)$, and the dashed ones are for
 the others specified in the vertical axes. The red line is for $a=0=W_\phi$.
 We have set $\lambda=+1$, $n=-2$,  $\omega=-0.85$,
$a_i=0.059$, $\phi(0)=0.88$, $\dot{\phi}(0)=-0.075$, and $\theta=1\times10^{-5}$.
 For greater visibility, we re-scaled the plots of $\dot{a}$, $\ddot{a}$ and $\ddot{\phi}$.
 We have used the Planck units.}
\label{phi-dif}
\end{figure}

 Moreover, for additional ICs as well as present parameters of our herein NC model, further figures are presented.
By depicting the behavior of $\dot{a}$, we have explained how the dynamics of
our NC model depend on the present parameters of the model, namely, $\omega$, $\theta$ and $n$.
In order to show that the model works well for other appropriate sets of the ICs
as well as parameters, let us here plot the time behavior of other physical
quantities in terms of different values of the present parameters, specially $n\leq0$, see, for instance, Figs. \ref{phi-dif}.

Up to now, employing a numerical analysis, we have shown
that the X-branch solutions in the JF yield an
accelerating scale factor for a short time, which can exit
successfully from this phase and subsequently turns to decelerate.
However, such results are not sufficient to consider them as a successful inflation.
In the next sections, we will investigate whether or not the
other required conditions for a successful kinetic inflation are satisfied.
%%%%%%%%%%%%%%%%%%%%%%%%%%%%%%%%%%%%%%%%%%%%%%%%%%%%%%%%%%%%%%%%%%%%%%%%%%%%%%%%%%%%%%%%%%%%%%%%%%%%

%%%%%%%%%%%%%%%%%%%%%%%%%%%%%%%%%%%%%%%%%%%%%%%%%%%%%%%%%%%%%%%%%%%%%%%%%%%%%%%%%%%%%%%%%%%%%%%%%%%%

%%%%%%%%%%%%%%%%%%%%%%%%%%%%%%%%%%%%%%%%%%%%%%%%%%%%%%%%%%%%%%%%%%%%%%%%%%%%%%%%%%%%%%%%%%%%%%%%%%%%

%%%%%%%%%%%%%%%%%%%%%%%%%%%%%%%%%%%%%%%%%%%%%%%%%%%%%%%%%%%%%%%%%%%%%%%%%%%%%%%%%%%%%%%%%%%%%%%%%%%%

\section{Horizon problem}
\label{Horizon}
As mentioned, obtaining an accelerating scale factor for the early times of the universe
is not solely a sufficient condition for getting a successful inflationary epoch.
More concretely, the most important problem with the standard cosmology, the
horizon problem, must be resolved in a successful inflationary scenario.
Moreover, in a successful inflation, the graceful exit problem must be solved.
 In the previous section, we have shown that our herein model
  does satisfy the latter. In this section, we would investigate the horizon problem.
In order to resolve the horizon problem, we should show that the extent of the
observable universe is encompassed by an inflated causally connected region.
In this respect, we will investigate both the nominal and sufficient conditions
associated to the obtained accelerating universe which is relevant for the causal physics of an inflation.

\subsection{Nominal condition}
In analogy with the standard models, the nominal condition can be written as
\begin{eqnarray}
D_{\rm H}\equiv d_{\rm H}^{\rm tot}-H^{-1}>0,\label{d-hor-1}
\end{eqnarray}
where
\begin{eqnarray}
d_{\rm H}^{\rm tot}\equiv a(t)\int\frac{d\bar{t}}{\bar{a}}.
\end{eqnarray}\label{d-hor-2}
We will show that in our model, $d_{\rm H}^{\rm tot}$ contributes two parts, namely, the commutative and NC parts  as
\begin{eqnarray}\label{d-hor-2}
d_{\rm H}^{\rm tot}= d_{\rm H}^{\rm com}+d_{\rm H}^{\rm nc},
\end{eqnarray}
such that when $\theta=0$, it reduces to the its corresponding commutative quantity.
In what follows, we will show that the nominal condition is completely
satisfied for the inflationary model obtained for the X-branch.

Following Ref. \cite{HTW94}, we employ (\ref{H2}) to obtain
\begin{eqnarray}
\left(H+\frac{\dot{\phi}}{2\phi}\right)^2=\frac{\xi^2}{4}\left(\frac{\dot{\phi}}{\phi}\right)^2,\label{hor-1}
\end{eqnarray}
which yields
\begin{eqnarray}
\frac{d{\rm ln}(\phi a^2)}{dt}=-\lambda\xi\left(\frac{\dot{\phi}}{\phi}\right).\label{hor-2}
\end{eqnarray}
Concerning our case where $\zeta(a)=a^n$, by employing (\ref{H-a}), (\ref{phidot-gen}) and
integrating (\ref{hor-2}) over $dt$, it is straightforward to show that
\begin{eqnarray}\label{d-hor-4}
d_{\rm H}^{\rm com}\!\!&=&\!\!\frac{\lambda (1-\delta) a^3 \phi}{b\xi}=\frac{\lambda a_i^3 (1-\delta) \phi^{3\chi+1}}{b\xi},\\\nonumber
\\
d_{\rm H}^{\rm nc}\!\!&=&\!\!-\frac{a_i^{n+1}\theta (2\chi+1)\phi^\chi}{6 b\xi^2[(n+1)\chi+2]}\int \phi^{n\chi+2}dt,\label{d-hor-5}
\end{eqnarray}
where we have
 included the integration constant $a_1^2 \phi_1$ in $\delta$, which was defined as
\begin{eqnarray}\label{delta}
\delta\equiv \frac{a_1^2 \phi_1}{a^2 \phi}.
\end{eqnarray}
Moreover, to calculate (\ref{d-hor-4}) and (\ref{d-hor-5}), we need to obtain $\phi$ in terms of the cosmic time, which
should be determined from the differential equation (\ref{phidot-gen}). We should note that
by considering an attractive gravity (where $\phi>0$ forever), then
$\delta$ always takes positive values. In this subsection,
%in contactwith the previous investigations \cite{Lev95,RM14},
let us assume $\delta=0$.
However, in the next subsection where the sufficient condition will be probed, we will
 proceed our calculations with nonzero $\delta$.

Therefore, by calculating the BD scalar field from (\ref{phidot-gen}), then substituting it
into relations (\ref{d-hor-4}) and (\ref{d-hor-5}), and finally obtaining the quantity $D_{\rm H}$ with the
assistance of numerical calculations, we can establish whether or not the nominal condition is satisfied in our NC model.
Because of the NC part $d_{\rm H}^{\rm nc}$, we expect that the horizon problem
can be solved in our NC model more appropriately than the generalized BD theory in which $\omega=\omega(\phi)$.
When $\theta=0$, then $d_{\rm H}^{\rm nc}$ vanishes and
we retrieve $D_{\rm H}=d_{\rm H}^{\rm com}$.

\begin{figure}
\centering\includegraphics[width=3.2in]{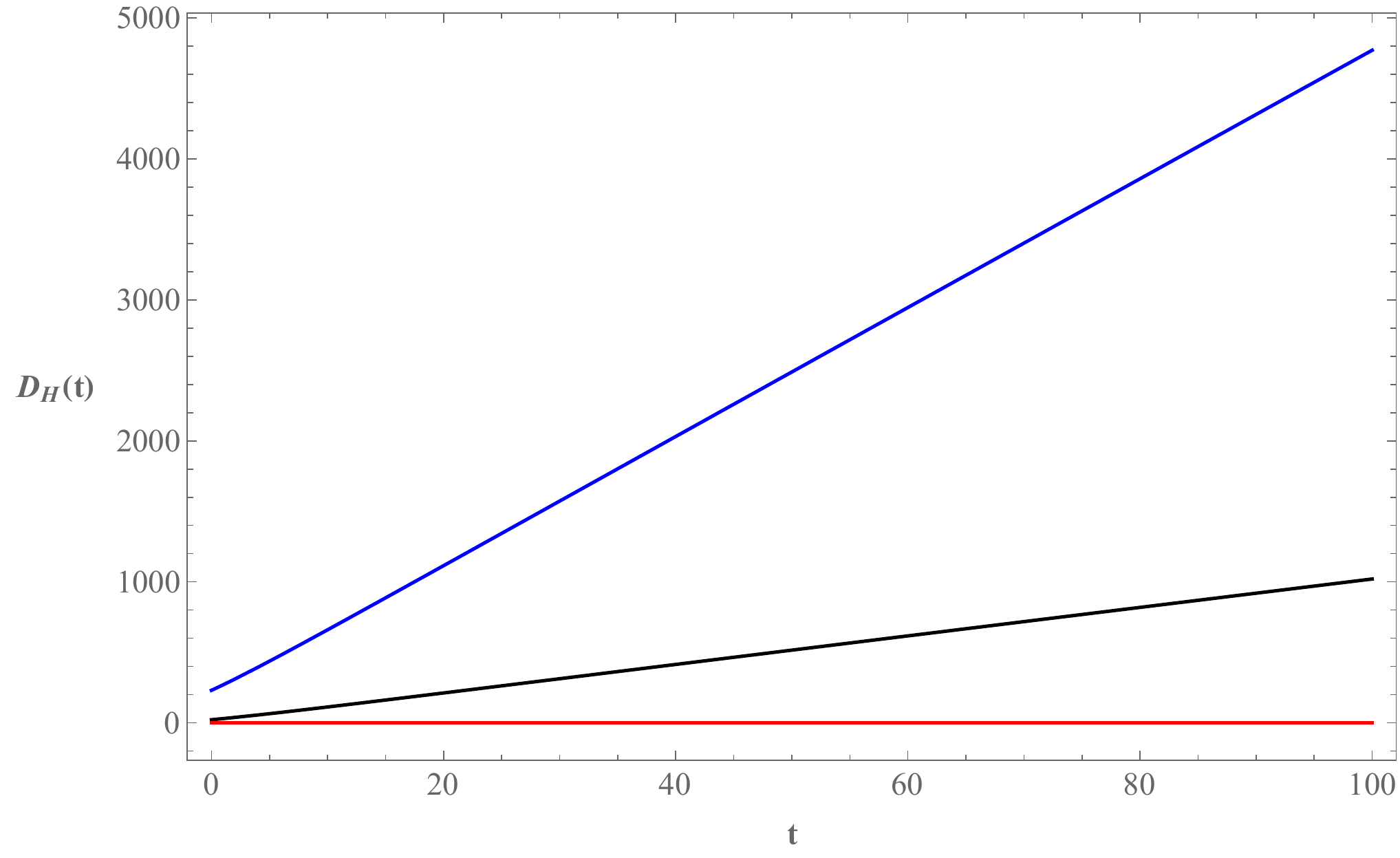}
\caption{ The time behavior of $D_{\rm H}$ for the NC case.
The black and blue curves have been plotted by taking the same values of the present parameters
and ICs used for figures \ref{phi-up} and \ref{phi-dif}, respectively. The red line is plotted for $D_{\rm H}=0$.
Moreover, we have set $b=|a_i^3\dot{\phi}(0)/10|$ and $\delta=0$.}
 \label{Dh}
\end{figure}

As mentioned, it is impossible to solve the differential equation (\ref{phidot-gen}), analytically, for the
 general cases. However, using numerical analysis, we have shown that for the NC
 case, the nominal condition is satisfied. For instance, for the upper solution ($\lambda=+1$), in which we
 have obtained an inflationary epoch for the early times and the zero
 acceleration for the scale factor at late times,
we have plotted the behavior of $D_{\rm H}$ in Fig. \ref{Dh}.

\subsection{Sufficient condition}
In analogy with the non-deformed case of generalized BD theory
in JF~\cite{Lev95-2}, we can admit the condition associated to the sufficient inflation as
\begin{eqnarray}\label{suf-hor-3}
 \frac{d_{\rm H_\star}^{\rm tot}}{a_\star}>\frac{1}{H_0a_0},
\end{eqnarray}
where $d_{\rm H_\star}^{\rm tot}$ and $a_\star$ denote the values of these quantities at an earlier time $t_\star$.
Moreover, the index $0$ denotes the present values of those quantities.
Obviously, in the particular case where $\theta=0$, from
relations (\ref{d-hor-2}), (\ref{d-hor-4}) and (\ref{d-hor-5}), it is easy to show that
the inequality (\ref{suf-hor-3}) reduces to its non-deformed counterpart.

Let us first calculate the left and right hand sides (lhs and rhs) of inequality (\ref{suf-hor-3}), separately, and then
 derive the sufficient condition. The lhs of (\ref{suf-hor-3}) is given by
\begin{eqnarray}\label{suf-4}
&{}& \frac{d_{\rm H_\star}^{\rm tot}}{a_\star}=-(a^2_\star \phi_\star)\left[
-\frac{\lambda (1-\delta)}{b\xi}
+\frac{a_i^{n-2}\theta (2\chi+1)\phi^{-(2\chi+1)}}
{6 b\xi^2[(n+1)\chi+2]}\int \phi^{n\chi+2}dt\right]_\star,
\end{eqnarray}
where the quantity inside brackets is evaluated at the specific
time $t_\star$ and we have used relations (\ref{H-a}), (\ref{d-hor-2}), (\ref{d-hor-4}) and (\ref{d-hor-5}).

For calculating the rhs of (\ref{suf-hor-3}), we will use the following steps employed in \cite{Lev95-2}:
(i) It is feasible to write the relation for the Hubble constant at present time as
\begin{equation}\label{h0}
H_0=\sqrt{\alpha_0}T_0^2,
\end{equation}
where $\alpha_0=\gamma(t_0)\eta_0=(8\pi/3)(\pi^2/30){g_\star}(t_0)\eta_0$; $\eta_0$ denotes
the ratio of the energy density in matter to that in
radiation at present time.
% whose value is given by $\eta_0\sim$ \cite{}.
(ii) We will use $t_{\rm end}$ to denote a specific time in which the inflation
was stopped and entropy was produced.
(iii) We will employ the relation $a_{\rm end}T_{\rm end}=a_0T_0$ which indicates an adiabatically evolution of the universe.
(iv) Concerning the heating mechanism, $T_{\rm end}$ can be related to the net available kinetic energy density $E_{\rm end}$ as
$T_{\rm end}=\epsilon E_{\rm end}$, where $\epsilon$ stands for the
efficiency whereby the kinetic energy density is converted to entropy~~\cite{Lev95-2}.
(v) Finally, let us assume that $E_{\rm end}$ is produced merely by the kinetic energy associated to the BD scalar field, namely,
$E_{\rm end}=(4\pi/3)a^3_{\rm end}\rho^{(\phi)}_{\rm end}$.
Consequently, by using assumptions (i)-(v) and employing relations (\ref{phidot-gen}) and (\ref{ro-3}), we obtain
\begin{eqnarray}\label{suf-5}
&{}&\frac{1}{H_0a_0}=\left[\frac{12 a^2_{\rm end}\phi_{\rm end} }{\epsilon (\omega-6\chi)\sqrt{\alpha_0}T_0}\right]\frac{1}{\left[b+\frac{a_i^{n+1}\theta(2\chi+1)\phi^{[(n+1)\chi+2]}}
{6\xi^2[(n+1)\chi+2]}\right]^2\Big|_{\rm end}}.
\end{eqnarray}

Substituting relations (\ref{suf-4}) and (\ref{suf-5}) into (\ref{suf-hor-3}) yields

\begin{eqnarray}\nonumber
 \frac{a_\star^2\phi_\star}{a_{\rm end}^2\phi_{\rm end}}
 &\gtrsim&
\left[\frac{12 \alpha_0^{-\frac{1}{2}} }{\epsilon (\omega-6\chi)T_0}\right]
 \frac{1}{\left|
-\frac{\lambda (1-\delta)}{b\xi}
+\frac{a_i^{n-2}\theta (2\chi+1)\phi^{-(2\chi+1)}}
{6 b\xi^2[(n+1)\chi+2]}\int \phi^{n\chi+2}dt\right|_\star}\\\nonumber
\\
&\times&\frac{1}{\left[b+\frac{a_i^{n+1}\theta(2\chi+1)\phi^{[(n+1)\chi+2]}}
{6\xi^2[(n+1)\chi+2]}\right]^2\Big|_{\rm end}}.\label{suf-hor-p}
 \end{eqnarray}
This is the modified (NC) version of the
sufficient condition associated to the inflation with respect to that obtained in
the generalized BD theory in \cite{Lev95-2} [where $\omega=\omega(\phi)$].

In \cite{Lev95-2}, in addition to the analytic analysis associated
to the sufficient conation in the JF, by choosing two different
suitable sets of the constants and the parameters of the model,
the following predictions have been demonstrated:
(i) For the first choice, it has been argued
that just the $D$-branch solutions (in which the quantity $a^2\phi$ always
decreases with cosmic time) can satisfy the sufficient condition. However, for
the present time, as $\phi$ must take constant value and the scale factor must
accelerate, therefore, the $D$-branch does not yield the
present expanding universe unless a branch change to be induced.
(ii) For the second choice, it has been demonstrated that the X-branch
(in which $a^2\phi$ may increase) can satisfy the sufficient condition if
one of the integration constants takes very large value.

It has been demonstrated that the sufficient condition associated to inflation
can be satisfied if the scale factor in the conformal EF does
accelerate~\cite{Lev95,Lev95-2}. Regarding this point and because of the
intricate construction of inequality (\ref{suf-hor-p}), let us skip the
intuitive analysis (where we should probe whether or not the sufficient condition, by
trying some sets of suitable values for the integration constants and the parameters,
is satisfied) and just produce our predictions by analyzing the
time behavior of the quantity $a^2\phi$. Let us also mention a relevant reason: it has been shown that obtaining an
 accelerating scale factor in the JF is not a sufficient condition
 for the occurrence of a corresponding inflation in the conformal EF.
 More concretely, another essential complementary condition must
  be satisfied: the following time dependent ratio must increase
  with the cosmic time (in the JF)~\cite{Faraoni.book}
\begin{eqnarray}\label{a-phi2}
\frac{a(t)}{l_{\rm Pl}(t)}=a(t)\sqrt{\phi(t)}\equiv a_{_{\rm RI}},
\end{eqnarray}
 where $l_{\rm Pl}(t)$ is the Planck length. Note that $a_{_{\rm RI}}$ is the scale factor associated to the EF (see section \ref{EF}).
If the requirements $\ddot{a}>0$ and $\ddot{a}_{_{\rm RI}}>0$ are satisfied in the
JF, then, following Ref.~\cite{Faraoni.book}, such an inflationary phase is called {\it real (actual) inflation.}
There are situations in which, although the scale factor accelerates, it cannot
increase faster than the $l_{\rm Pl}(t)$, see for instance \cite{C98, Lev95-2}.
In \cite{C98}, the pre-big bang cosmology~\cite{LWC00} has been studied in
the BD theory for the specific value $\omega=-1$, where the BD theory
corresponds to the low energy limit of some string theories.

\begin{figure}
\centering\includegraphics[width=2.5in]{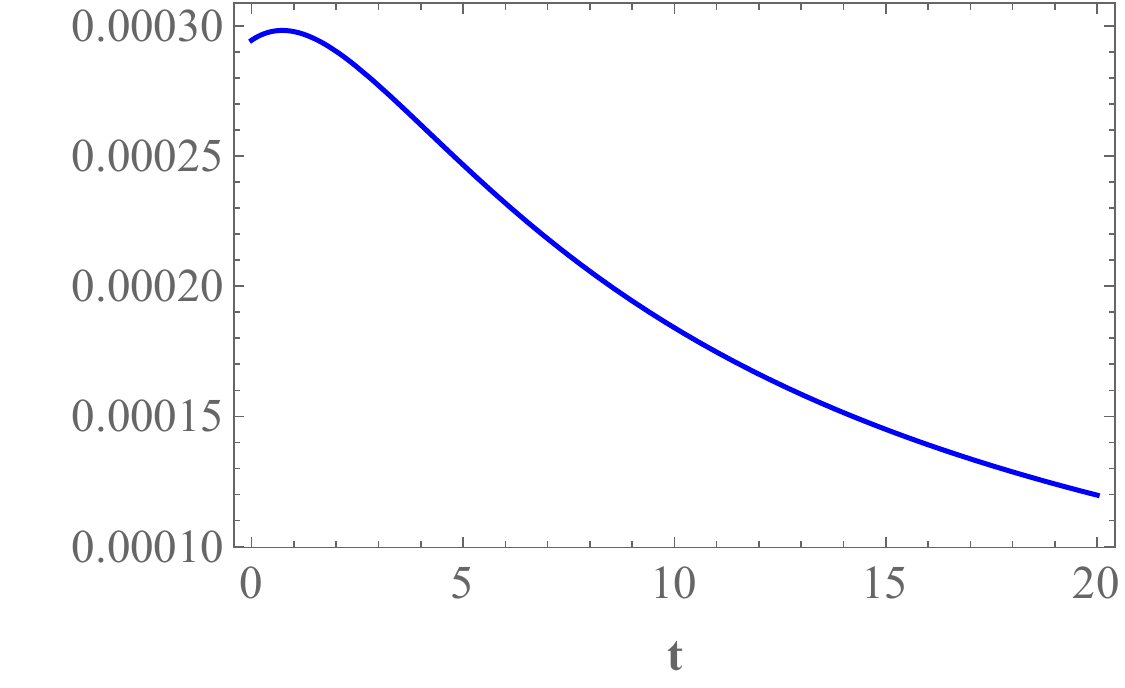}
\hspace{5mm}
\centering\includegraphics[width=2.5in]{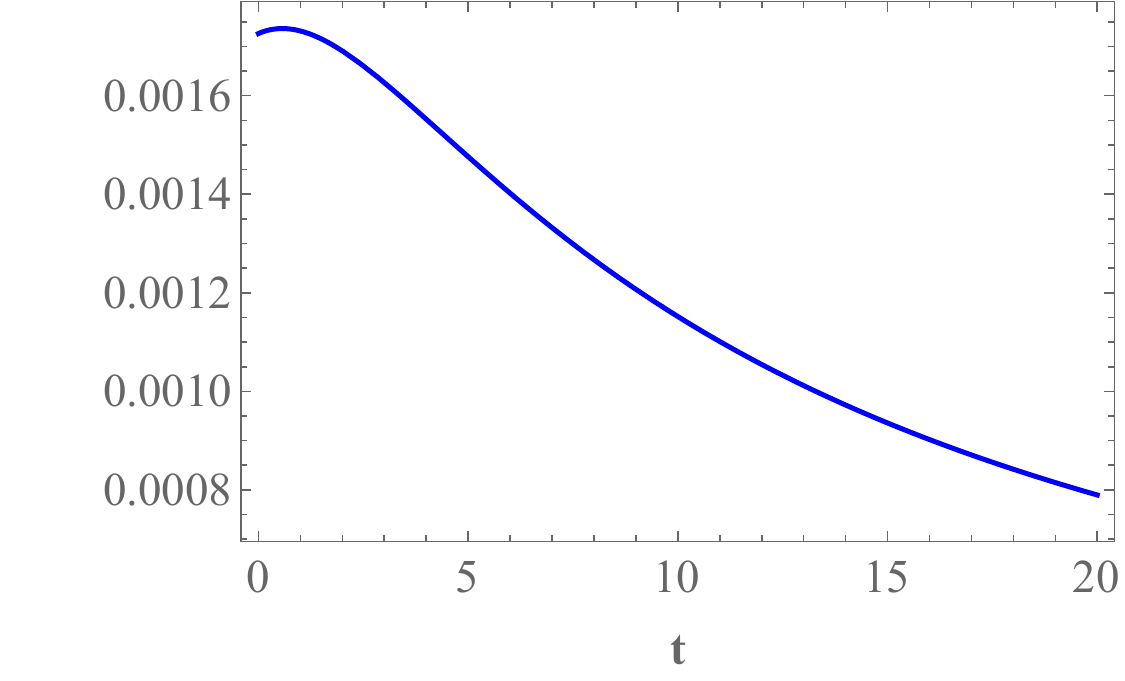}
\centering\includegraphics[width=2.5in]{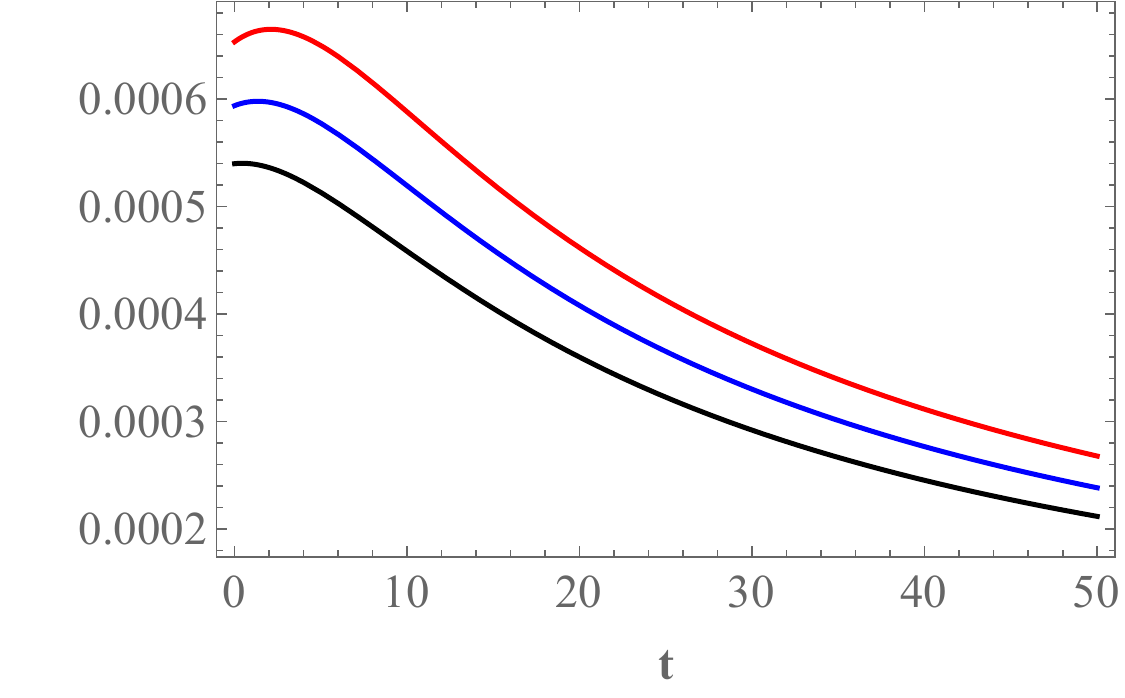}
\hspace{5mm}
\centering\includegraphics[width=2.5in]{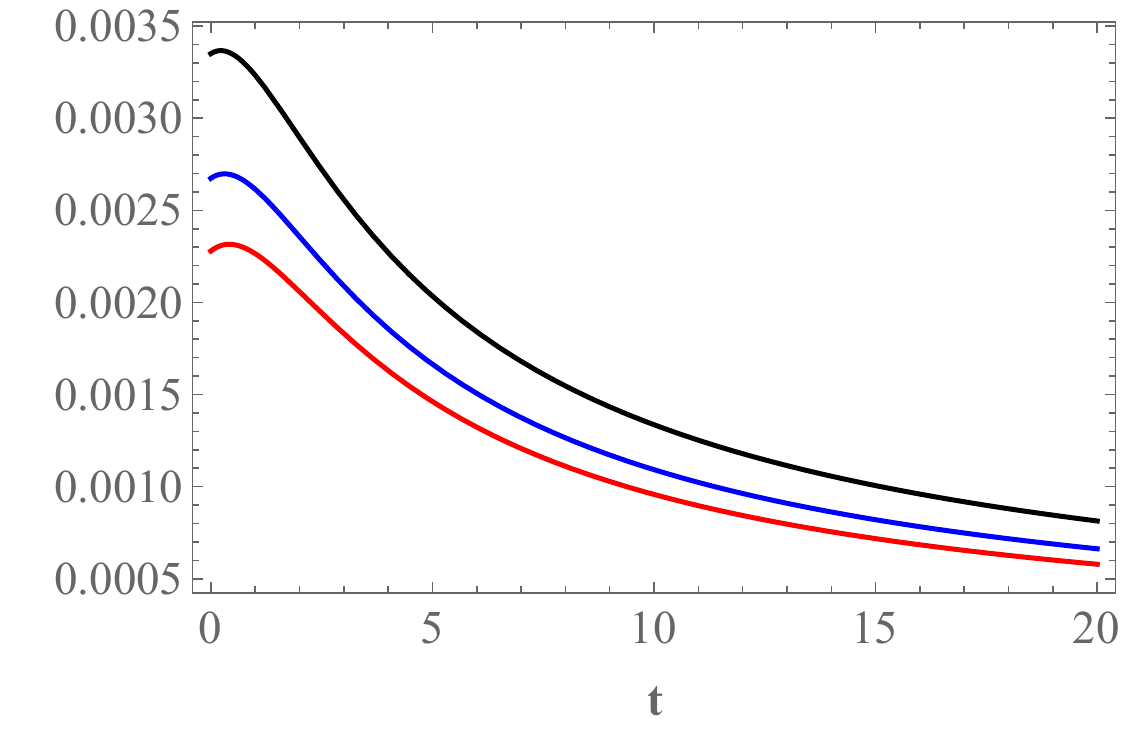}
\caption{ The behavior of $\frac{d}{dt}(a\sqrt{\phi})$  against cosmic time.
For the {\it upper left panel} and {\it upper right panel}, we have chosen
the same ICs and parameters of Figs. \ref{phi-up}-\ref{ro-p-up} and Fig. \ref{phi-dif}, respectively.
Lower left panel: we have employed the same ICs used in the left panel of Fig.
  \ref{adot-dif-par-up}. More concretely, we have set
  $\theta=8\times10^{-5}$ (black curve), $\theta=9\times10^{-5}$ (blue curve),
$1\times10^{-4}$ (red curve), $\omega=-1.1$, $n=0$, $a_i=0.05$, $\phi(0)=2$ and $\dot{\phi}(0)=-0.11$.
 Lower right panel: we have employed the same ICs used in the right panel of Fig.
  \ref{adot-dif-par-up}, namely, $\omega= -0.8$ (black curve), $\omega= -0.9$
(blue curve), $\omega= -1$ (red curve), $a_i= 0.05$, $n=0$, $\lambda=1$,
$\theta=1\times10^{-4}$, $\phi(0)=3.8$ and $\dot{\phi}(0)=-0.98$.
For greater visibility, plots of the lower panels have been re-scaled.
We have used the Planck units.}
 \label{RI}
\end{figure}

 For the resulted
inflation associated to the upper solution in section~\ref{Vacuum-NC-BD}, again, by means of
numerical analysis, we have shown that the quantity $a_{_{\rm RI}}(t)$ not only always increases
 with the cosmic time, but also accelerates at some time prior to the present time.
 In other words, for the upper solution in which we have taken very
 small positive values for $\theta$ and the BD coupling parameter is restricted to $-3/2<\omega<0$,
 the sufficient condition is always satisfied. For instance, in Fig. \ref{RI}, we
 have plotted the quantity $\frac{d}{dt}(a\sqrt{\phi})$ versus cosmic time. It is important to
 note that for the specific case where $\omega=-1$, we have also obtained a real
 inflation, see Fig. \ref{RI}. Moreover, to show that our NC model works well
 for other appropriate ICs, we will also provide other plots for
 the quantity $\frac{d}{dt}(a\sqrt{\phi})$ in the \ref{RI}.

\section{Cosmological dynamics in deformed phase space Brans-Dicke theory}
\label{dynamical}
In this section, let us first compare the evolutionary equation for the
scale factor associated to our herein kinetic
inflation, arisen from deformed phase space BD setting,
with that obtained in ``$R^2$'' (``Starobinsky'') inflationary model \cite{S80}.
Subsequently, we will proceed our discussion to a dynamical analysis and plot the phase
 plane diagrams.
 % with the same range of initial conditions used in the previous section.
% These analysis do confirm the resulted
 %inflationary epoch with graceful exit.

Employing equations (\ref{a2dot}) and (\ref{H-a}), it is straightforward to show that
the equation for the scale factor associated to $\zeta(a)=a^n$ case can be written as
%\begin{eqnarray}\nonumber
%\frac{\dot{a}\dddot{a}}{a^2}&+&(4-n)\frac{\dot{a}^2\ddot{a}}{a^3}-\left(\frac{\ddot{a}}{a}\right)^2\\\nonumber
%\\\nonumber
%\!\!&+&\!\!\Big\{(n-4)-\frac{2}{3(1+\lambda \xi)^3}\left[(n-4)+(n-2)\lambda \xi\right]\\
%\!\!&\times&\!\!\left[6\xi(\xi+\lambda)-\omega\right]\Big\}\left(\frac{\dot{a}}{a}\right)^4=0,
%\end{eqnarray}\label{a-evol-Gen-NC}
\begin{eqnarray}\label{a-evol-Gen-NC}
\frac{\dot{a}\dddot{a}}{a^2}+(4-n)\frac{\dot{a}^2\ddot{a}}{a^3}-\left(\frac{\ddot{a}}{a}\right)^2+A\left(\frac{\dot{a}}{a}\right)^4=0
\end{eqnarray}
 where
 \begin{eqnarray}
A\!&\equiv&\!(n-4)-\frac{2}{3(1+\lambda \xi)^3}\left[(n-4)+(n-2)\lambda \xi\right]
\left[6\xi(\xi+\lambda)-\omega\right].\label{A}
\end{eqnarray}
In the commutative case, where $\theta=0$, equation (\ref{a-evol-Gen-NC}) reduces to
\begin{eqnarray}
\frac{\ddot{a}}{a}+\Bigg[\frac{2\xi(\xi+\lambda)}{(1+\lambda \xi)^2}\Bigg]\left(\frac{\dot{a}}{a}\right)^2=0.
\end{eqnarray}\label{a-evol-com-case1}

Moreover, it is worthwhile to employ a
few appropriate transformations introduced in \cite{S80}.
First, let us set $u=a^2\dot{a}^2$. Then, equation (\ref{a-evol-Gen-NC}) transforms into
\begin{eqnarray}
\frac{uu''}{a^2}&+&(2-n)\frac{uu'}{a^3}-\frac{u'^2}{2a^2}
+2\left[A+n-2\right]\frac{u^2}{a^4}=0,\label{a-evol-2}
\end{eqnarray}
where a prime stands for a derivative with respect to the scale factor $a$.
From equations (\ref{a-evol-2}), we obtain a partial de Sitter solution:
\begin{eqnarray}
u(a)=H_1^2a^4, \hspace{10mm}a(t)=a_1{\rm exp}(H_1t),
\end{eqnarray}
(where $a_1$ and $H_1$ are integration constants) provided that
\begin{equation}\label{a-evol-6}
\omega=\left\{
 \begin{array}{c}
-\frac{4}{3} \hspace{33mm} {\rm for}\hspace{5mm} \lambda=-1\\\\
\frac{6(3-n)}{(n-2)^2} \hspace{2mm} ({\rm where} \hspace{2mm} n\neq2)
\hspace{5mm} {\rm for}\hspace{5mm} \lambda=\pm1.
 \end{array}\right.
\end{equation}
Second, by letting $g=u^{\frac{3}{4}}$ and $z=(12)^{-\frac{3}{4}}a^3$, (\ref{a-evol-2}) is represented as
\begin{eqnarray}
\frac{d^2g}{dz^2}&-&\frac{1}{3g}\left(\frac{dg}{dz}\right)^2-\left(\frac{n-4}{3z}\right)\frac{dg}{dz}
+\left[A+n-2\right]\frac{g}{6z^2}=0.\label{a-evol-3}
\end{eqnarray}
\\
Finally, by substituting
\begin{eqnarray}\label{xy}
g=zx, \hspace{10mm} y=z\frac{dx}{dz}
\end{eqnarray}
into (\ref{a-evol-3}), it yields
\begin{eqnarray}
\frac{dy}{dx}=\frac{y}{3x}-\frac{\left[A+4-n\right]x}{6y}+\frac{n-5}{3}.\label{a-evol-4}
\end{eqnarray}
From equations (\ref{a-evol-Gen-NC}), (\ref{a-evol-2}), (\ref{a-evol-3}) and (\ref{a-evol-4}), it
is seen that, disregarding the absence of the term $H^2$ as
well as different multiplications associated to some terms, these
equations bear much resemblance to the corresponding ones obtained in \cite{S80}.
 The price we had to pay to obtain such a correspondence between
 our herein model and that obtained in \cite{S80} was to remove the explicit presence of the
 NC parameter in the equations derived above.
Further complicated investigations are required to derive the role
of the NC parameter together with BD coupling parameter (as well as other
parameters and integration constants) via the employed transformations to construct a physical connection from our
model to the elements used in \cite{S80}. This investigation is not the scope of the present work.

Let us move to the second stage of this section in which we can see the
explicit presence of the NC parameter in our dynamical analysis.
Taking a constant BD coupling parameter and employing $\zeta(a)=a^n$
and (\ref{H-a}), equation (\ref{a2dot}) can be rewritten as
\begin{equation}
\dot{H}+c_1(\omega)\: H^2+\theta\:c_2(\omega)\: a^{\left(n-2+\frac{1}{\chi}\right)}H=0,
\label{H-tilde}
\end{equation}
where
\begin{align}
c_1(\omega) &= \dfrac{1}{3\chi^2} \left(3\chi^2+\omega-3\chi \right), \\
c_2(\omega) &= \dfrac{\omega -3\chi}{18\:a_i^{\frac{1}{\chi}}\chi\:\xi^2}.
\end{align}
Let $y=\dot{a}$, then equation (\ref{H-tilde}) transforms to
\begin{equation}
\dfrac{dy}{da}=\Big[1-c_1(\omega) \Big]a^{-1}y-\theta c_2(\omega)a^{\left(n-2+\frac{1}{\chi}\right)}\: .
\label{y-dyn}
\end{equation}
\begin{figure}
\centering\includegraphics[width=2.4in]{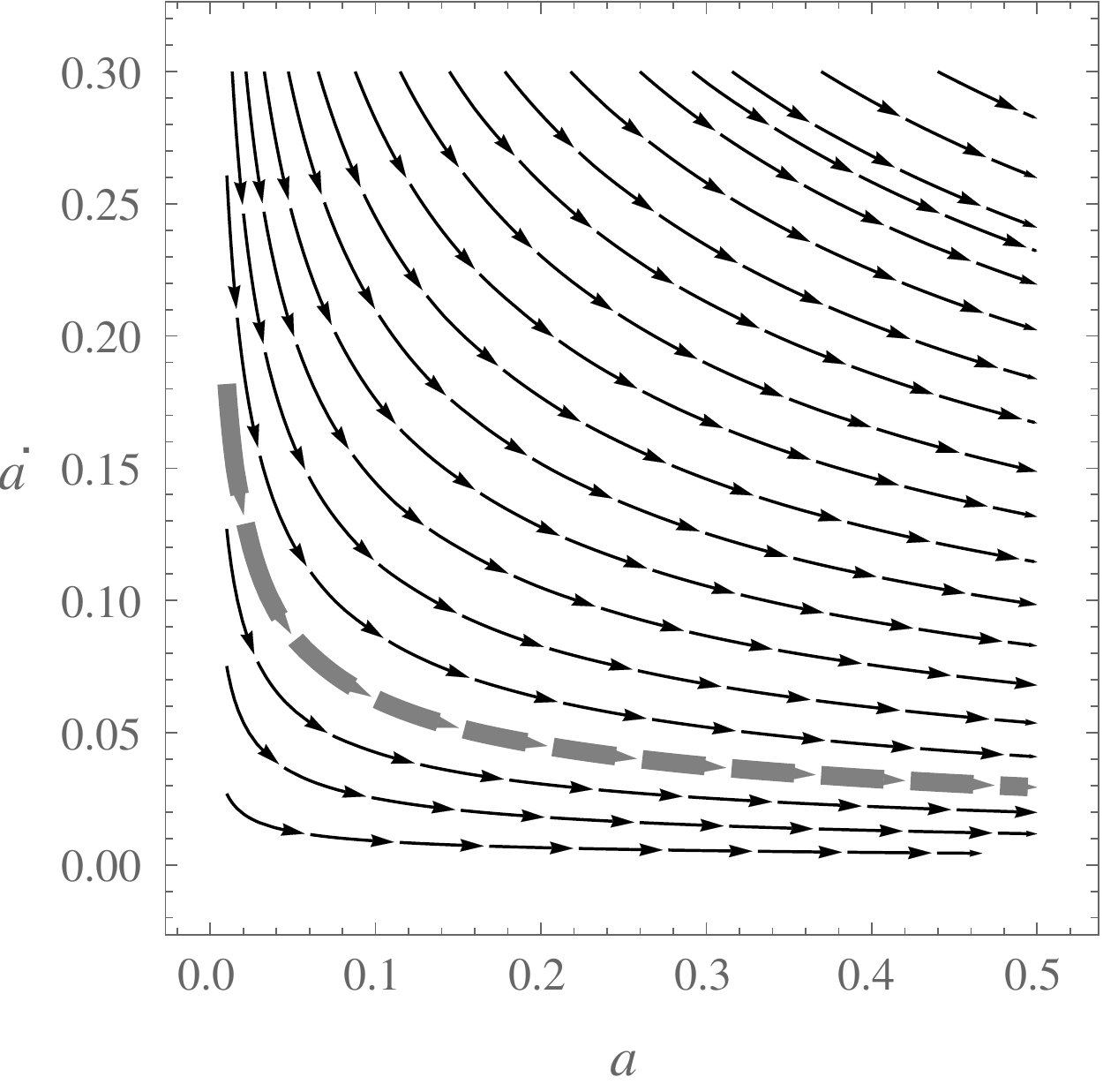}
\hspace{5mm}
\centering\includegraphics[width=2.4in]{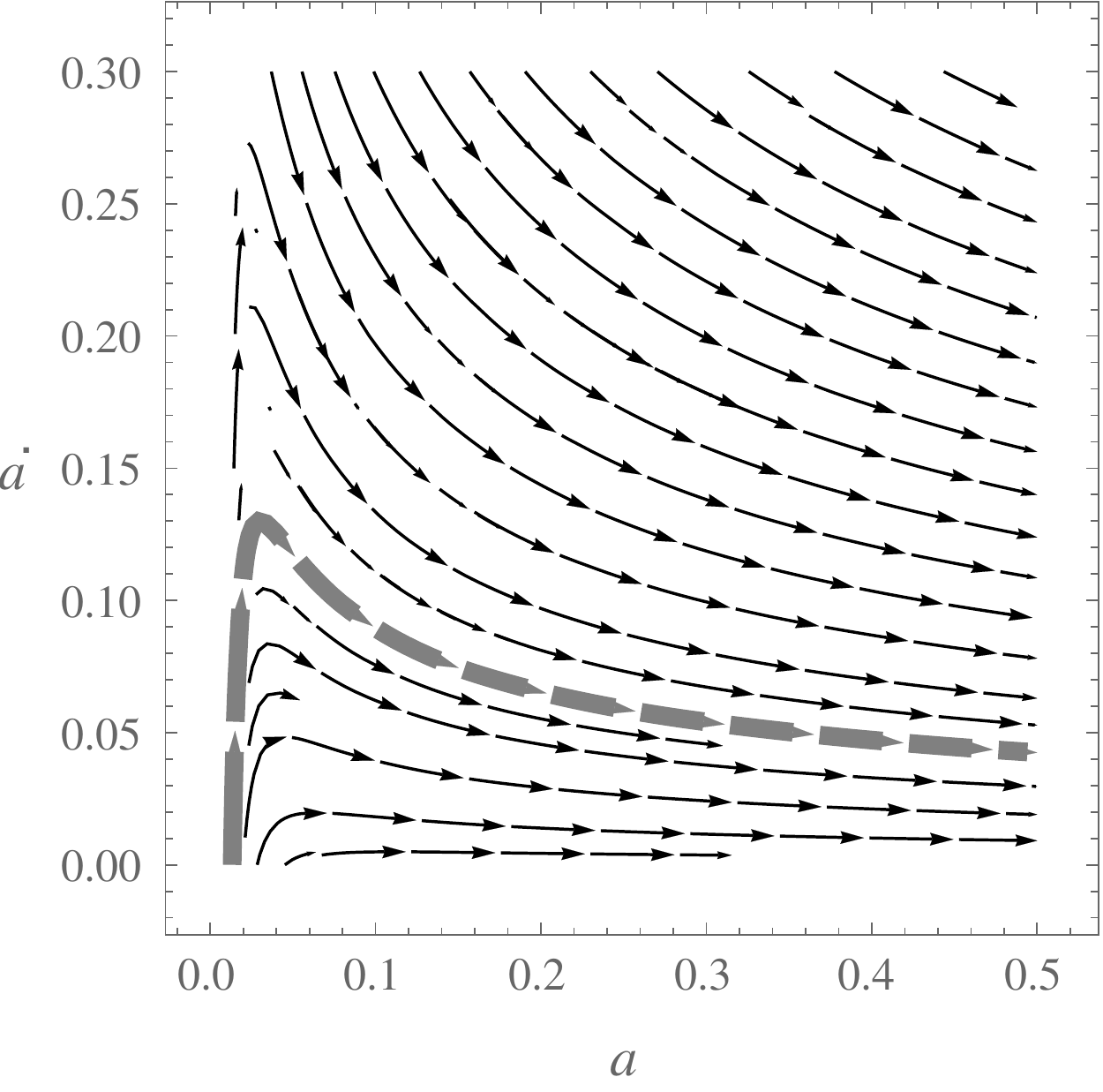}
\caption{Phase space portrait for Eq. (\ref{y-dyn}) for the X-branch ($\lambda=+1$).
The left and right panels are associated to the commutative
case ($\theta=0$) and noncommutative case ($\theta=0.001$), respectively.
We have set $\omega=-1.35$ and $a_i=0.07$ for both of the cases.}
\label{PhaseP}
\end{figure}
This equation enables us to illustrate the difference between the commutative and
noncommutative cases. In the left panel of Fig. \ref{PhaseP}, we plotted
the phase portrait of equation (\ref{y-dyn}) when $\theta=0$ (non-deformed case).
We observe that all the solutions $(\dot{a},\:a)$ start with large and always decreasing
values of $\dot{a}$. When the noncommutative parameter $\theta$ is switched on (see the right panel)
the situation changes and all the solutions start with very small values for $\dot{a}$,
which increases until a maximum is reached. Afterwards, $\dot{a}$ begins its decreasing phase.
We should note that the particular
solution in Fig. \ref{a-up} corresponds to one of the trajectories depicted in Fig. \ref{PhaseP},
thus showing that a vast range of solutions allow an acceleration stage.
This behavior is in accordance with the inflationary epoch
belonging graceful exit, which was already obtained for the NC case
associated to the X-branch in section \ref{Vacuum-NC-BD}. Finally, we should note that it is straightforward to show that
similar trajectories as shown in Fig. \ref{PhaseP} can be also plotted for other sets of proper ICs and parameters of the model.

%introduced in \cite{S80}.
%First, let us set $u=a^2\dot{a}^2$. Then, equation (\ref{a-evol-Gen-NC}) transforms into
%\begin{eqnarray}\nonumber
%\frac{uu''}{a^2}&+&(2-n)\frac{uu'}{a^3}-\frac{u'^2}{2a^2}\\
%&+&2\left[A(\omega)+n-2)\right]\frac{u^2}{a^4}=0,\label{a-evol-2}
%\end{eqnarray}
%where a prime stands for a derivative with respect to the scale factor $a$.
%From equations (\ref{a-evol-2}), we obtain partial de Sitter solution:
%\begin{eqnarray}
%u(a)=H_0^2a^4, \hspace{10mm}a(t)=a_i{\rm exp}(H_0t),
%\end{eqnarray}
%where $a_i$ is an integration constant, provided that
%\begin{equation}\label{a-evol-6}
%\omega=\left\{
% \begin{array}{c}
%-\frac{4}{3} \hspace{33mm} {\rm for}\hspace{5mm} \lambda=-1\\\\
%\frac{6(3-n)}{(n-2)^2} \hspace{2mm} ({\rm where} \hspace{2mm} n\neq2)
%\hspace{5mm} {\rm for}\hspace{5mm} \lambda=\pm1,
 %\end{array}\right.
%\end{equation}
%Second, by letting $g=u^{\frac{3}{4}}$ and $z=(12)^{-\frac{3}{4}}a^3$, (\ref{a-evol-2}) is represented as
%\begin{eqnarray}\nonumber
%\frac{d^2g}{dz^2}&-&\frac{1}{3g}\left(\frac{dg}{dz}\right)^2-\left(\frac{n-4}{3z}\right)\frac{dg}{dz}
%\\
%&+&\left[A(\omega)+n-2)\right]\frac{g}{6z^2}=0,\label{a-evol-3}
%\end{eqnarray}
%\\
%Finally, by substituting
%\begin{eqnarray}\label{xy}
%g=zx, \hspace{10mm} y=z\frac{dx}{dz}
%\end{eqnarray}
%into (\ref{a-evol-3}), it yields
%\begin{eqnarray}
%\frac{dy}{dx}=\frac{y}{3x}-\frac{\left[A(\omega)+4-n)\right]x}{6y}+\frac{n-5}{3}.
%\end{eqnarray}\label{a-evol-4}
\section{Einstein Frame}
\label{EF}
For obtaining the new set of dynamical variables $(\tilde{\gamma},\tilde{\phi})$ associated
to the EF\rlap,\footnote{It should be emphasized that the discrepancies/equivalencies
of the different conformal frames have been
widely discussed, see, for instance, \cite{Lev95-2,Faraoni.book} and references
therein, and further inquiry as to the details of such notifications will not be in the scope of the present work.}
we should use the conformal transformation for the metric
together with redefinition for the scalar field as~\cite{Faraoni.book}
\begin{equation}
\gamma_{\mu\nu}=\Omega^2 \tilde{\gamma}_{\mu\nu}, \hspace{10mm} \Omega=\phi^{-\frac{1}{2}},
\label{con-1}
\end{equation}
\begin{equation}
\tilde{\phi}(\phi)=\sqrt{\frac{3}{16 \pi}}\xi {\rm ln}\phi,
\label{con-2}
\end{equation}\\
 where, again, we have used the Planck units, $\phi\neq0$ and the BD coupling parameter must be
restricted as $\omega\geq-3/2$ to guarantee to get real values for $\tilde{\phi}$.
Moreover, in order to compare the results of our model with those obtained
in the non-deformed case, let us employ the procedure using the same
coordinate transformation employed in \cite{Lev95-2}
\begin{equation}
dt=\Omega d\tilde{t}, \hspace{10mm} {\rm and} \hspace{10mm}a=\Omega \tilde{a}.
\label{con-3}
\end{equation}
From transformations (\ref{con-1})-(\ref{con-3}), the metric in the EF
 is written as the spatially flat FLRW metric as \cite{Lev95-2}
\begin{equation}\label{metric-EF}
d\tilde{s}^{2}=-d\tilde{t}^2+\tilde{a}^{2}\left(dx^2+dy^2+dz^2\right),
\end{equation}
where the scale factor in the EF is related to that
in the JF as $\tilde{a}= a\phi^{1/2}$ and, again, the lapse
function has been set equal to unity.

It is straightforward to show that the evolutionary equations
associated to the scale factor and the scalar field
correspond to the EF can be written as\footnote{In order to obtain the field
equations in the EF, we have transformed the corresponding equations
associated to the JF, under the conformal transformations.
Equivalently, we can derive the herein equations of motion (in EF) by starting from the
 conformal EF action. Therefore, we must construct the corresponding deformed Poisson bracket
 in the EF by employing the the required transformation and respecting the dimensional
 analysis provided that, again, the rhs depends linearly on the NC parameter.}

\begin{eqnarray}\label{EF-H-1}
 \tilde{H}^2\!\!\!&=&\!\!\!\frac{8\pi}{3}\tilde{\rho}_{\tilde{\phi}},\\\nonumber
\\
2\frac{d^2\tilde{a}}{d\tilde{t}^2}
\!\!\!&+&\!\!\!3\tilde{H}^2=-8\pi\tilde{p}_{\tilde{\phi}},\\\nonumber
\\
\frac{d^2\tilde{\phi}}{d\tilde{t}^2}\!\!\!&+&\!\!\!3\tilde{H}\frac{d\tilde{\phi}}{d\tilde{t}}=-\left(\frac{\theta \tilde{a}^{n-2}\tilde{H}}{4\sqrt{3\pi}\xi}\right)
 {\rm exp}\left[\frac{2\sqrt{3\pi}(3-n)\tilde{\phi}}{3\xi}\right],\label{EF-phi}
\end{eqnarray}
where $\tilde{H}$ is the Hubble parameter in the EF and
\begin{eqnarray}\label{ro-EF}
\tilde{\rho}_{\tilde{\phi}}&=&\frac{1}{2}\left(\frac{d\tilde{\phi}}{d\tilde{t}}\right)^2,\\\nonumber
\\
\label{p-EF}
\tilde{p}_{\tilde{\phi}}&=&\frac{1}{2}\left(\frac{d\tilde{\phi}}{d\tilde{t}}\right)^2+\frac{\sqrt{3\pi}}{36 \pi}\left(\frac{\theta \tilde{a}^{n-2}}{\xi}\right)
 {\rm exp}\left[\frac{2\sqrt{3\pi}(3-n)\tilde{\phi}}{3\xi}\right]\frac{d\tilde{\phi}}{d\tilde{t}}
\end{eqnarray}
are the energy density and the pressure associated to the scalar field $\tilde{\phi}$, respectively. The EoS parameter for the EF is defined as
$\tilde{W}_{\tilde{\phi}}=\tilde{p}_{\tilde{\phi}}/\tilde{\rho}_{\tilde{\phi}}$.

%, and
%\begin{eqnarray}\label{k}
%\kappa\equiv\frac{2\sqrt{3\pi}(3-n)}{3\xi M_{\rm Pl}}.
%\end{eqnarray}
From (\ref{EF-H}), we can obtain the relation between the scale factor and the scalar field as
\begin{eqnarray}\label{EF-H}
\tilde{H}=-\lambda\sqrt{\frac{4\pi}{3}}\left(\frac{d\tilde{\phi}}{d\tilde{t}}\right),
\hspace{10mm}\tilde{a}(\tilde{t})=\tilde{a}_i {\rm exp}\left[-\lambda\sqrt{\frac{4\pi}{3}}\tilde{\phi}(\tilde{t})\right],
\end{eqnarray}
where $\tilde{a}_i$ is an integration constant.

Therefore, equation (\ref{EF-phi}) is easily integrated and yields
\begin{eqnarray}
\frac{d\tilde{\phi}}{d\tilde{t}}=\frac{1}{\tilde{a}^3}
\left[-\tilde{C}+\frac{\theta \tilde{A}}{\tilde{B}}{\rm exp}\left(\tilde{B}\tilde{\phi}\right)\right]
,\label{phi-first-int}
\end{eqnarray}
where
\begin{eqnarray}\nonumber
\tilde{A}\equiv \frac{\lambda \tilde{a}_i^{n+1}}{6\xi}, \hspace{3mm}
\tilde{B}\equiv-\frac{2\sqrt{3\pi}}{3\xi}\left[(n+1)\lambda\xi+(n-3)\right]
\end{eqnarray}
and $\tilde{C}$ is an arbitrary integration constant.
By substituting $\tilde{a}$ from (\ref{EF-H}) into equation (\ref{phi-first-int}), we obtain
\begin{eqnarray}
\frac{d\tilde{\phi}}{d\tilde{t}}=\tilde{a}_i^3{\rm exp}
\left[-2\lambda\sqrt{3\pi}\tilde{\phi}(\tilde{t})\right]\left[-\tilde{C}+\frac{\theta \tilde{A}}{\tilde{B}}{\rm exp}\left(\tilde{B}\tilde{\phi}\right)\right],
\label{phi-first-int-2}
\end{eqnarray}
 which is a first order differential equation only in terms of scalar field $\tilde{\phi}$.
  However, for the general case, the corresponding solutions lead us to the Hypergeometric functions.

 %which is related to $c$ [appeared in equation (\ref{con-2})] via (\ref{phidot-gen}) as
%\begin{eqnarray}\nonumber
%\tilde{c}=-\frac{3c\kappa \xi^2}{8\pi(3-n)}.\label{c-tild}
%\end{eqnarray}
In the particular case where $\theta=0$, equations (\ref{EF-H}), (\ref{EF-phi}),
(\ref{phi-first-int}) and (\ref{phi-first-int-2}) reduce to the their
corresponding standard counterparts in the EF~\cite{Lev95-2,C98}.
As the solutions and the pertinent problems associated to the standard case
 have been widely studied in~\cite{Lev95-2,C98}, let us forbear from reviewing them.
 However, let us just review two important points \cite{Lev95}. (i) Since only $\dot{\phi}<0$ corresponds to inflation in the JF,
therefore, let us focus on the case where $d\tilde{\phi}/d\tilde{t}<0$.
In this case, both constants $\tilde{C}$ and $b$ (which is introduced in equation (\ref{phidot-1})), which
 are related as $\tilde{C}=b\sqrt{\frac{3}{16 \pi}}$, should take positive values.
Consequently, from (\ref{EF-H}), we find that only the upper sign ($\lambda=+1$) yields an expanding scale factor of the universe in the EF which corresponds to our upper (X-branch) solution in the JF. (ii) By using relations
 (\ref{EF-phi}), (\ref{EF-H}) and (\ref{phi-first-int}), we can easily obtain
 \begin{eqnarray}\label{a-sec}
\frac{d^2\tilde{a}}{d\tilde{t}^2}=-\frac{8\pi \tilde{C}^2}{3\tilde{a}^5},\\\nonumber
\end{eqnarray}
which implies that at all times $d^2\tilde{a}/d\tilde{t}^2<0$, namely, the
scale factor of X-branch solution in the EF expands in a decelerating rate.

 In the previous sections, we have shown
 that, by choosing very small positive values of the NC parameter together with
 appropriate sets of ICs, it is feasible to obtain successful
 kinetic inflation in both the frames.
 In order to re-corroborate the consequences associated to the
 EF, by employing numerical approaches directly arisen from
 the equations of the EF, we can easily shown that an inflationary phase is obtained, in which the
 corresponding nominal and sufficient conditions are satisfied.
 In this respect, we have used the advantage of the following approach.
  By substituting $\tilde{H}$ and $\tilde{a}(\tilde{t})$ from (\ref{EF-H}) into (\ref{EF-phi}), we obtain
\begin{eqnarray}\label{EF-phi-2}
\frac{d^2\tilde{\phi}}{d\tilde{t}^2}
-\sqrt{12\pi}\lambda\left(\frac{d\tilde{\phi}}{d\tilde{t}}\right)^2=
\lambda\theta\left(\frac{\tilde{a}_i^{n-2}}{6\xi}\right)
{\rm exp}\left[\lambda\sqrt{\frac{4\pi}{3}}\left(2-n+\frac{(3-n)\lambda}
{\xi}\right)\tilde{\phi}\right]\frac{d\tilde{\phi}}{d\tilde{t}}.
\end{eqnarray}

\begin{figure}
\centering\includegraphics[width=2.5in]{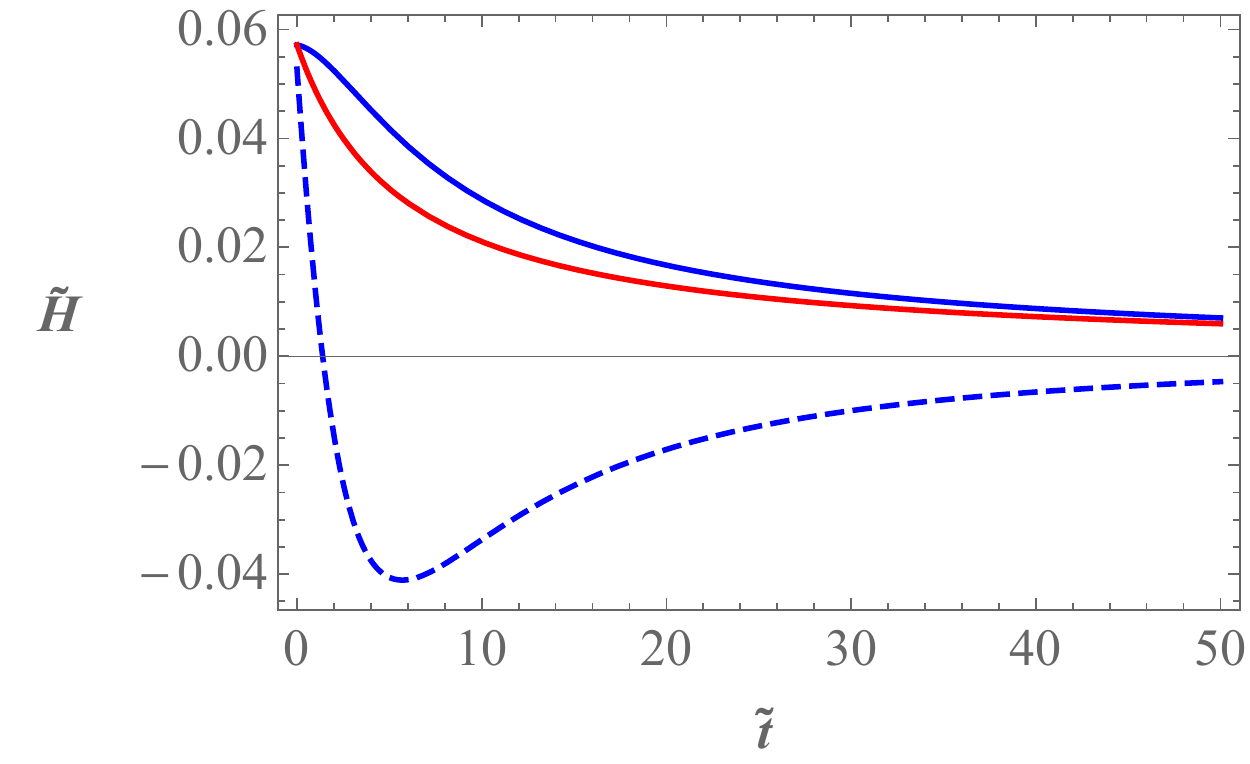}
\hspace{5mm}
\centering\includegraphics[width=2.5in]{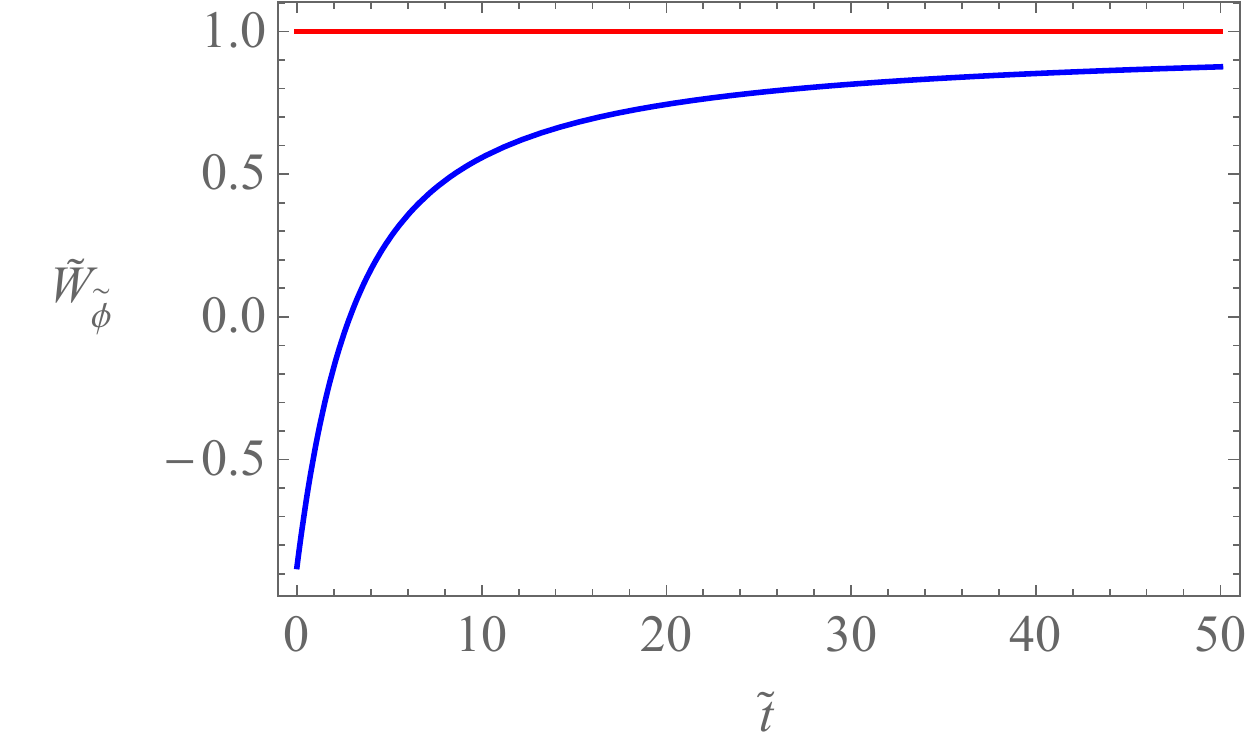}
\caption{ In the left and right panels, the time behaviors of the $\tilde{H}(\tilde{t})$
and $\tilde{W}_{\tilde{\phi}}(\tilde{t})$ have been depicted, respectively, for the commutative case
(red curves) and NC case (blue curves). The dashed blue curve shows the time behavior of $\frac{d^2\tilde{a}}{d \tilde{t}^2}$ for the NC case.
We have set $\lambda=+1$, $\xi \approx 0.98$, $\tilde{a}_i=8.5$, $\tilde{\phi}(0)=5$, $\frac{d\tilde{\phi}(0)}{d \tilde{t}}=-0.14$,
$n=0$ and $\theta=0.0001$ (for the NC case).
 To depict this figure, for convenience, we used the
 geometrical units where $\hbar=c=8\pi G=1$.}
\label{EF-H-EoS}
\end{figure}

 By employing reasonable sets of ICs and pertinent ranges of the
present parameters of the model, numerical solutions for (\ref{EF-phi-2})
together with (\ref{ro-EF})-(\ref{EF-H}) have shown that
 the consequences associated to the EF confirm our previous achievements, especially those of section \ref{Horizon}.
 More concretely, our numerical endeavors have shown that, similar to the JF, by restricting
the freedom in choosing the ranges of ICs
such that $-1<\tilde{W}_{\tilde{\phi}}<1$, taking very small positive values of NC
parameter and the BD coupling constant from the interval $-3/2<\omega<0$, we can still
obtain an appropriate inflationary phase for the X-branch at early times.
Let us abstain from re-depicting the time behavior of the all
quantities and merely focus on the time
behavior of the EoS parameter $\tilde{W}_{\tilde{\phi}}$ and the Hubble parameter, see, for instance, Fig. \ref{EF-H-EoS}.

In what follows, we would like to derive the sufficient
condition (\ref{suf-hor-3}) in the EF and see that how it is modified with respect to the non-deformed case.
Similar to the method employed in~\cite{Lev95-2}, it is
straightforward to show that inequality (\ref{suf-hor-3}) associated to our NC model in the EF can also be written as
\begin{eqnarray}\label{EF-suf-1}
 \frac{\tilde{d}_{\rm H_\star}^{\rm tot}}{\tilde{a}_\star}>\frac{1}{\tilde{H}_0\tilde{a}_0},
\end{eqnarray}
where using relations (\ref{EF-phi}) and (\ref{EF-H}), it is easy to show that
\begin{eqnarray}\label{EF-suf-2}
 \tilde{d}_{\rm H}^{\rm tot}(\tilde{t})&=&
 \tilde{a} \int_{\tilde{t}_i}^{\tilde{t}}\frac{d\tilde{t}'}{\tilde{a}(\tilde{t}')}=\frac{\sqrt{3\pi}\lambda \tilde{a}^3|1-\delta|}{4\pi \tilde{C}}
 +\frac{\theta \tilde{A}\tilde{a}}{\tilde{B}\tilde{C}}\int_{\tilde{t}_i}^{\tilde{t}}\frac{{\rm exp}[\tilde{B}\tilde{\phi}(\tilde{t}')]d\tilde{t}'}{\tilde{a}(\tilde{t}')},
 \end{eqnarray}
 where $\delta=(a_1^2 \phi_1)/(a^2\phi)=\tilde{a}_1^2/\tilde{a}^2$. Reemploying (\ref{EF-phi}) and (\ref{EF-H}) and applying the same
  approximation used in~\cite{Lev95-2}, it is easy to show that the sufficient condition
  associated to the EF in our NC model is given by
  \begin{eqnarray}
&{}& \left(\frac{d\tilde{a}}{d\tilde{t}}\Big|_\star\right)^{-1}-\frac{\theta \tilde{A}}{\tilde{B}\tilde{C}}
 \left[\left(\frac{d\tilde{a}}{d\tilde{t}}\right)^{-1}{\rm exp}\left(\tilde{B}\tilde{\phi}\right)\right]_\star \nonumber
 \\\nonumber
 \\
&+&\frac{\theta \tilde{A}}{\tilde{a}_i\tilde{B}\tilde{C}}\left\{\int_{\tilde{t}_i}^{\tilde{t}}{\rm exp}\left[\left(\tilde{B}+\frac{2\lambda \sqrt{3\pi}}{3}\right)\tilde{\phi}(\tilde{t}')\right]d\tilde{t}'\right\}_\star
\gtrsim  \left(\frac{d\tilde{a}}{d\tilde{t}}\Big|_0\right)^{-1}.\label{EF-suf-3}
 \end{eqnarray}
 In order to evaluate this inequality we must find $\tilde{\phi}(\tilde{t})$ by
  solving the differential equation (\ref{phi-first-int-2}).

 The above inequality in the non-deformed case where $\theta=0$, reduces to the
 corresponding counterpart obtained in ~\cite{Lev95-2}, namely,
  \begin{eqnarray}\label{EF-suf-4}
\frac{d\tilde{a}}{d\tilde{t}}\Big|_0\gtrsim \frac{d\tilde{a}}{d\tilde{t}}\Big|_\star.
 \end{eqnarray}

  It has been demonstrated that for satisfying the sufficient condition,
  the scale factor associated to EF must accelerate at some time prior to the present time \cite{HTW94,Lev95-2}.
 However, in the standard BD theory it is not possible to get an accelerating scale factor in EF.
  However, as we have already shown in the previous sections, the time behaviors of scale factors in both the JF
 and EF (namely, $\tilde{a}\propto a \phi^{1/2}$) for the X-branch solutions, have indicated that
  our NC model yields inflationary epoch possessing graceful exit in both of the
  frames. Therefore, checking inequality (\ref{EF-suf-3}) in this section will be just a rehash of the same issue.

  Before closing our investigations in the EF, let us
  represent the equations of motion associated to the EF in the conformally flat background where the conformal time $\eta$ is introduced as
  \begin{eqnarray}\label{conf-time}
d\tilde{t}= \tilde{a}(\eta)d\eta.
 \end{eqnarray}
It is straightforward to show that equations (\ref{EF-H}) and (\ref{EF-phi}), in
terms of the conformal time, can be combined as
 \begin{eqnarray}\label{con-eq}
-\frac{8}{3\xi}\sqrt{\frac{\pi }{3}}\tilde{a}\left[\frac{3}{4\pi}\frac{d^2\tilde{a}}{d\eta^2} +\tilde{a}\left(\frac{d\tilde{\phi}}{d\eta}\right)^2\right]\!+\!\frac{d\left(\tilde{a}^2\frac{d\tilde{\phi}}{d\eta}\right)}{d\eta}\!=\!\frac{\theta \tilde{a}_i^{n+1}(3\lambda\xi+4) }{18 \xi \tilde{B}}\frac{d\left[{\rm exp} \left(\tilde{B}\tilde{\phi}\right)\right]}{d\eta},
 \end{eqnarray}
where we have used (\ref{EF-H}). In the non-deformed case
where $\theta=0$, equation (\ref{con-eq}) is substantially simplified as
\begin{eqnarray}\label{con-eq-2}
\frac{d^2\tilde{\phi}}{d\eta^2}-\sqrt{12\pi}\lambda
\left(\frac{d\tilde{\phi}}{d\eta}\right)^2-\frac{d{\rm ln}\tilde{a}}{d\eta}\frac{d\tilde{\phi}}{d\eta}=0.
 \end{eqnarray}

 With respect to equation (19)
of \cite{GOZ99}, under particular conditions with $\theta\neq0$, which can make the rhs of
equation (\ref{con-eq}) equal to zero, we recover that equation.
It might be expected, by means of numerical analysis, at the level of equations of motion, we can get a sensible
correspondence between our herein NC model with the quantum BD
cosmological model proposed in \cite{GOZ99,GK00}, in which the conformally flat background setting
for the spinor case in the EF has been established.
%In this case, the solutions for equations (\ref{EF-H}), (\ref{EF-phi}) are
%\begin{eqnarray}
%\tilde{a}(\tilde{t})\!\!\!&=&\!\!\!\left[\tilde{a}_i^3\pm\sqrt{12\pi}\tilde{c}M_{\rm %Pl}^{-1}(\tilde{t}-\tilde{t}_i)\right]^{\frac{1}{3}},\label{a-con-EF-sol}\\\nonumber
%\\
%\tilde{\phi}(\tilde{t})\!\!\!&=&\!\!\!\tilde{\phi}_i\pm\frac{M_{\rm Pl}}{\sqrt{12\pi}}\rm{ln}\left[\tilde{a}_i^3\pm\sqrt{12\pi}\tilde{c}M_{\rm %Pl}^{-1}(\tilde{t}-\tilde{t}_i)\right]\!\!,\label{phi-con-EF-sol}
%\end{eqnarray}
%where $\tilde{t}_i$, $\tilde{a}_i$ and $\tilde{\phi}_i$ are integration constants.

\section{Conclusions}
\label{conclusion}
In this paper,
we introduced a dynamical deformation\footnote{The special case of the
deformation (\ref{NC-Poisson}) has been previously proposed in Ref. \cite{RFK11}, based upon dimensional
 analysis; the Jacobi identity being satisfied.} in the phase space associated to the momenta of the
 spatially flat FLRW scale factor and the BD scalar field.
  Using the Hamiltonian formalism, after some computations, we have derived the modified equations,
 such that when the NC parameter vanishes, we recover the
 equations correspond to the non-deformed case.
 In particular, from applying the proposed deformed Poisson bracket, we found that the Hamiltonian
 constraint is not modified, but other equations contains extra terms
 explicitly depending on the NC parameter (cf. Section \ref{NC-BDT}).

For the sake of comparison with the standard case, we have
presented (see Section \ref{Vacuum-NC-BD}) the exact solutions
associated to the non-deformed case and described the properties of the solutions.
Moreover, we have pointed that the pre and post big-bang solutions obtained in our
herein paper does obviously include those obtained for
the particular case of the BD theory with $\omega=-1$ in Ref. \cite{C98}.

Subsequently, we proceeded towards our specific NC setting and took a constant
BD coupling parameter, set up to work with a general function
$\zeta(a)$ [in the rhs of the deformed Poisson bracket
in (\ref{NC-Poisson})] as well as with the particular case where $\zeta(a)=a^n$.
 We have shown how the equations of motion associated to the scale factor, BD scalar field, energy density and pressure
  (associated to the BD scalar field) only depend on the dynamics of the BD scalar field.
  We have further obtained general conditions under which the scale factor can describe an
  accelerating (a decelerating) universe and the constraints determining appropriate energy conditions.

After obtaining particular solutions associated to the NC model, as it was not
possible to find exact analytic solutions associated to the complicated coupled differential
equations associated to the general NC case, we focused on a numerical analysis.
Concerning the D-branch solutions (see Section \ref{Vacuum-NC-BD}), we have shown that it is
possible to get a kinetic inflationary epoch for the universe.
%However, these solutions are the same as those
%obtained for the BD cosmology by taking either
%constant or varying BD coupling parameter have encountered with the problems.
Despite employing a highly diverse set of ICs in the
corresponding numerical endeavors, we could not find any suitable
condition\footnote{We should note that we have merely restricted ourselves to the natural and reasonable sets of ICs.} which can produce an inflationary epoch possessing satisfactory exit from inflation.
More concretely, our herein NC kinetic inflation associated to
the D-branch, similar to those obtained in \cite{Lev95-2}, suffers from the graceful exit problem.
However, motivated from constructing another distinct NC model in the context of the BD theory \cite{RM14}, where we had previously
overcome this problem for the inflationary model associated to D-branch, we have
preferred to focus on the X-branch solutions, which in the commutative cases,
even with varying $\omega$, are not immune from the problems \cite{Lev95-2}.

 Let us be more precise, so to clarify the above previous sentence.
   By employing the allowed ranges of the ICs for the X-branch, we have
 shown that the evolution of the universe for both the early
 and the late times fits better with the current observational data
 than those obtained in the non-deformed models.
 More concretely, concerning the X-branch solutions, in contrary to the models constructed in
 the context of the BD theory (even with varying $\omega$), we have obtained solutions with the following features.
 In the Jordan frame (JF), there is no big bang singularity for the inflationary
 phase, occurring at early times of the universe. After a short time, the accelerating
 phase is replaced by a decelerating epoch which can be assigned to the radiation era.
 Such a successful graceful exit is emerges from the NC effects, without needing any branch-change process.
 It should be emphasized that in the standard and/or generalized
 BD theory, it is not possible to get a kinetic inflation with graceful exit.
 In contrast to the commutative cases, for late times, the scale factor tends
 to take a constant value (such a consequence is in agreement with those obtained in the
 previous NC models, see, for instance, \cite{RM14, RFK11}, and references therein), which
 might be supposed to signify the effects
 of the NC effects on the very large scales (coarse grained explanation by the quantum gravity effects).
 In such a universe,
  the energy density and  pressure associated to the BD scalar
  field always take positive and negative values, respectively.
 Moreover, these solutions have been probed by employing a variety of different ICs.

 Then, we have focused on the horizon problem (see Section \ref{Horizon}) by considering the nominal as well as sufficient conditions.
 We have shown that both these conditions are satisfied for our herein X-branch solutions.

 Concerning the cosmological dynamics (see Section \ref{dynamical}), we have
 focused on two different stages. In the first stage, in order to compare our NC
 model with the well-known Starobinsky model at the level of equations of motion, not only we have
 constructed the second order differential equation associated to
 deformed equations for the scale factor, but also we have produced
 a useful discussion.
 It has been demonstrated that there might be a close resemblance between our model and the Starobinsky
 model, at least in the level of the equations of motion.
These equations indicate that it might be possible to establish a physical relation between our NC model with the Starobinsky
 model via deriving a relation which can connect the NC as well as BD coupling
 parameter to constants associated to quantum field contributions.
 However, as we had to remove the NC parameter for drawing such a
 comparison, constructing such relations is a very complicated issue.
 In the second stage, we have constructed another representation for the
 equation of the scale factor, in which the NC parameter is present.
 We have shown that the phase plane diagrams associated to the
  X-branch solutions confirm the corresponding numerical results.

  Furthermore, we have derived the equations of our herein NC model in the Einstein frame (EF).
Similar to the standard case, the Hubble parameter is proportional to the
corresponding time derivative of the scalar field associated to the EF. However, the
wave equation which yields the scalar field has an extra term which comes from the
dynamics of the NC portion. From a EF view point, this extra term appropriately assists
 to obtain a real inflation for the X-branch for our model. For $\theta=0$, we
  can recover the corresponding exact solutions associated to the non-deformed case.
  Moreover, we have obtained the sufficient condition for inflation in the EF.
  It has been further pointed that satisfying the required
  conditions associated to real inflation in the JF
  is equivalent to not only getting inflation in the EF with a graceful
   exit, but also satisfying the sufficient conditions in both the frames.
   %we have abstained from repeating the equivalent analysis.
   Subsequently, we have established the EF field equations in the
   conformally flat background, which indicate that
   we might obtain a sensible correspondence between our herein NC framework
   with the Einstein representation of the quantum BD setting
   constructed in the conformally flat background, where the scalar field coupled to the Dirac spinor matter.

In summary, in the herein paper, we have mainly focused on the X-branch solutions
in the absence of any scalar potential.
 %and we studied the
%effects of the quantum/semiclassical consequences and compared them with the classical as well as other quantum inflationary models.
 By extending the present model into more generalized frameworks,
% we expect that not only the pertinent problems can be solved
in the presence of NC effects,
  but with more degrees of freedom, we will get further solutions with interesting features. In fact:\\
%(i) We have only presented a brief discussions regarding the pre and
% post big-bang scenario for the particular case where $\theta=0$. Such a model can be extrapolated
 %from the current trends to involve the NC effects to produce more
 %interesting solutions for solving the problems associated to the standard models.
(i) Exploring the freedom in choosing the arbitrary function $\zeta(a)$ (not as power-law form), it may
extend our ability to solve the problems associated to X
and D branches, simultaneously. However, performing such a nontrivial
complicated calculations was not in the scope of the present work.\\

 (ii) Employing the herein model but taking a varying BD
 coupling parameter, it is expected to get solutions in which it may be feasible to overcome the mentioned problem,
  associated to the D-branch solutions.
 Our expectation lies in that due to having NC effects, with varying $\omega$,
 it would be possible to generalize the discussions regarding a bouncing universe already presented in \cite{Lev95}.\\

 (iii) In the presence of the ordinary matter and considering instead a
 non-flat FLRW space time (but still assuming vanishing
 scalar potential), due to the presence of the
 %semiclassical/quantum effects arisen from our herein
 NC effects, we will have more degrees of freedom with
 respect to the corresponding standard models (see for instance, \cite{LF93} in the case of varying $\omega$).
 %in obtaining solutions and resolving the probable pertinent problems.
 Moreover, by taking
 constant values for $\omega$ and considering the equation of state associated
 to the false vacuum, we can construct a NC model of extended inflation \cite{MJ84,LS89}. Furthermore,
in the more general cases, by assuming non-vanishing
 scalar potential and the ordinary matter
 fields, with varying, and even with constant values of the BD
 coupling parameter, it will be worthwhile to compare the results of the
 generalised NC inflationary model with those obtained in the standard cases, see for instance, \cite{BC91}.\\

 (iv) As the FLRW setting is homogeneous and isotropic, it has been claimed
 that investigating the horizon problem in such framework is an apparent paradox \cite{HTW94}.
  Respecting this argument, as well as in order to obtain
 observables for compatibility checking of our model with the detailed observational data from Planck \cite{Pl1,Pl2},
  we should construct a cosmological perturbation setup for our herein NC model. However, the lack of a
  definite Lagrangian corresponding to the modified set of
  equations of motion (\ref{H2})-(\ref{phi2dot}) is an obstruction
  for applying the linear perturbations theory. Nevertheless, it
  would be very interesting to see how the Mukhanov-Sasaki equation
  would be modified with the introduction of a NC parameter.
  This would enable to get explicitly the power spectrum of the
  vacuum fluctuations and all the relevant cosmological observational parameters.
 This is a very intriguing and difficult task that we leave for a forthcoming paper.\\

 (v) It has been believed that the most appropriate approach to address
that whether or not the cosmological problems (such as horizon and flatness problems) can be
solved is to rewrite the field equations of the model in terms of number of
e-foldings (instead of cosmic time) and obtaining the amount of inflation.
It is important to emphasize that we should
distinguish the different definitions of the e-fold number, which have been introduced as
either $\bar{N}$ or $N$ in the literature (for a detailed presentation, see,
for instance, \cite{LPB94, C16}, and references therein).
Let us be more precise.
The first condition for inflation is $\ddot{a}>0$, which can be rewritten as $d[(a H)^{-1}]/dt<0$.
Moreover, in a viable inflation model, in which the horizon and flatness problems can be
solved, the reduction in the size of $1/aH$ rather than of $1/a$ must be considered.
Furthermore, concerning the generation of perturbations, the relation of the wavenumber to inverse
Hubble radius $aH$ should be considered \cite{LPB94}.
Consequently, the e-fold number is defined as $\bar{N}\equiv {\rm ln} [(a H)_{\rm f}/(a H)_{\rm i}]$ (in Ref. \cite{C16}, $\bar{N}$ was called physical e-fold number) where the indices $i$ and $f$ refer to the initial and end of the inflation, respectively.
For some specific simple models, for instance, extreme slow roll limits potential
slow roll and Hubble slow roll (in which the Hubble parameter
is constant compared to the scale factor during inflation),
 $\bar{N}$ and $N\equiv {\rm ln} (a_{\rm f}/a_{\rm i})$ coincide \cite{LPB94, C16}.
 The other important point is that for the models whose Hubble
 parameter varies sufficiently during inflation, then $\bar{N}$ and $N$ differ significantly, and
 therefore, instead of the standard slow-roll approaches, we should
 establish the Hamilton-Jacobi formalism (for detailed review, see, for
 instance \cite{LLKCA97,C14} and references therein) to describe the corresponding
 generic properties \cite{C16}. Therefore, with respect to the above mentioned comments, let us focus on
 our herein NC model and argue why we have not established
 our herein NC discussions in terms of the (physical) number of e-foldings.
 (I) In one hand, we have employed the nominal and
sufficient requirements in terms of the cosmic time [cf inequalities (\ref{d-hor-1}) and (\ref{suf-hor-3})] and we
have shown that the horizon problem is satisfied in both the JF and EF settings.
We should note that it is claimed that the inequality (\ref{suf-hor-3}) is the strongest requirement (in terms of the cosmic time) to
address the satisfaction of the horizon problem \cite{HTW94,Lev95-2}. Moreover, it has been demonstrated that
if the horizon problem can be solved then there is no requirement to
solve the other cosmological problems, specially the flatness problem \cite{HTW94}.
It should be noted that the condition (\ref{suf-hor-3}) is equivalent to the
condition yielding a definition for the (physical) e-fold number.
(II) On the other hand, even disregarding the argument (I), our numerical
endeavors have shown that for any specific appropriate set of ICs, the
 Hubble parameter of our model varies during inflationary period, see, for instance, Figs. \ref{a-up} and \ref{phi-dif}.
 Therefore, to reconstruct our NC model versus the number of e-fold, we must employ the more general definition $\bar{N}$, and therefore
 %In addition, we have not employed any kind of conventional (slow-roll) approximations for our model. Therefore,
 concerning analyzing the dynamics of inflation, we thought worthwhile to employ instead
 the Hamilton-Jacobi formalism for our NC model, which, due to the
 presence of the NC sectors, was not a straightforward procedure and we subsequently left it out of the scope of the preset study.

\section{Acknowledgements}
 We express our sincere gratitude to anonymous referee for the careful
reading of our paper and the raised fruitful comments.
We would like to thank Shahram Jalalzadeh for his valuable comments.
S. M. M. Rasouli is grateful for the support of
Grant No. SFRH/BPD/82479/2011 from the Portuguese
Agency Funda\c c\~{a}o para a Ci\^encia e Tecnologia. PVM is grateful for DAMTP hospitality during his
sabbatical.
This research work was supported by Grant
No. UID/MAT/00212/2019 and COST Action CA15117 (CANTATA).

\bibliographystyle{utphys}

\end{document}